\documentclass[preprint,prl, preprintnumbers,superscriptaddress,amsmath,amssymb]{revtex4-2}

\usepackage[dvipdfmx]{graphicx}
\usepackage{colortbl}
\usepackage{mathrsfs}
\usepackage{lineno}
\usepackage{multirow}
\usepackage{siunitx}

\usepackage{bm} 
\usepackage{ulem}
\usepackage{comment}
\usepackage{amsmath}
\usepackage{autobreak}

\renewcommand{\Re}{\operatorname{Re}}
\renewcommand{\Im}{\operatorname{Im}}

\makeatletter
\def\btt#1{\texttt{\@backslashchar#1}}%
\DeclareRobustCommand\bblash{\btt{\@backslashchar}}%
\makeatother

\begin{document}


\title{Reply to  ``Comment on ``Reconsidering the nonlinear emergent inductance: time-varying Joule heating and its impact on the AC electrical response'''' by Yokouchi \textit{et al.}} 

\author{Soju Furuta}
\affiliation{Department of Physics, Institute of Science Tokyo, Tokyo 152-8551, Japan}

\author{Wataru Koshibae}
\affiliation{RIKEN Center for Emergent Matter Science (CEMS), Wako 351-0198, Japan}

\author{Taka-hisa Arima}
\affiliation{RIKEN Center for Emergent Matter Science (CEMS), Wako 351-0198, Japan}
\affiliation{Department of Advanced Materials Science, University of Tokyo, Kashiwa 277-8561, Japan}

\author{Fumitaka Kagawa}
\email{kagawa@phys.sci.isct.ac.jp}
\affiliation{Department of Physics, Institute of Science Tokyo, Tokyo 152-8551, Japan}
\affiliation{RIKEN Center for Emergent Matter Science (CEMS), Wako 351-0198, Japan}

\date{\today}
\begin{abstract}
This is a response to the comments [arXiv:2407.15682 and Phys.~Rev.~B \textbf{111}, 146401 (2025)] by Yokouchi \textit{et al}.~on our paper in Phys.~Rev.~B \textbf{110}, 174402 (2024). In this Reply, we note that (i) their arguments lack a discussion of whether the overall characteristics of the observed nonlinear impedance, including its magnitude and unphysical negative inductance interpretation, can be explained by the emergent induction scenario, whereas the Joule heating model can, (ii) they incorrectly refer to the Joule heating model, and (iii) their new data in the Comment are also quantitatively explained by the Joule heating model. These findings suggest that, contrary to the opinion by Yokouchi \textit{et al.}, the overall behavior of the observed impedance is irrelevant to emergent induction.
\end{abstract}

\maketitle

\newpage


\subsection{Fundamental problem of emergent induction with negative inductance}

In Ref.~\cite{YokouchiNatureA}, Yokouchi \textit{et al}.~measured $j$-nonlinear impedance, $\Delta Z(j, \omega) = Z(j, \omega) - Z(j \approx 0, \omega)$, where $j$ and $\omega$ denote the current density and angular frequency respectively, of a noncollinear magnet Gd$_{3}$Ru$_{4}$Al$_{12}$. They observed negative $\Im \Delta Z(j, \omega)$ proportional to $\omega$. From this observation, they concluded that they found an emergent induction, an electromotive force of spin-dynamics origin induced by a change in electric current, and that the sign of the inductance was negative \cite{YokouchiNatureA}. However, their argument raises a question regarding the validity of the emergent-induction interpretation because the inductance of a passive element should be positive \cite{JacksonA}. This question has been raised by our group \cite{Furuta1A} and, more recently, by another group \cite{KoreaA}. $\omega$-linear negative $\Im Z$ in a conducting two-terminal device is normally regarded as capacitive, i.e., a parallel circuit of a capacitance $C$ and resistance $R$, but not as negatively inductive, i.e., a series circuit of a negative inductance $L$ and $R$. Specifically, in an electric circuit, $\omega$-linear $\Im Z$ can be either positive or negative, but only a positive value is allowed for the inductance defined by $V = L \frac{{\rm d}I}{{\rm d}t}$, where $V$ and $I$ are the counterelectromotive force and current, respectively. As demonstrated in our paper \cite{FurutaPRBA}, even if negative inductance was assumed in the $j$-nonlinear regime, it could never reproduce $\omega$-linear negative $\Im \Delta Z$, as observed in their experiment. Yokouchi \textit{et al.}~have commented on our paper \cite{FurutaPRBA}, maintaining the scenario of emergent induction of a negative sign, without any discussion on the discrepancy with the fact that $\omega$-linear negative $\Im Z$ is not equivalent to an induction of a negative sign.

In our paper \cite{FurutaPRBA}, we argue that the overall characteristics of the observed $\Delta Z$ are explained even quantitatively by considering time-varying Joule heating, implying that, contrary to the opinion of Yokouchi \textit{et al.}~\cite{CommentA}, the observed impedance responses are irrelevant to emergent induction. In the following, we note that (i) the arguments of Yokouchi \textit{et al.}~lack a discussion of whether the overall characteristics of the observed $\Delta Z(j, \omega)$, including its magnitude, can be explained by the emergent induction scenario, whereas the Joule heating model can, (ii) they incorrectly refer to the Joule heating model, and (iii) their new data in the Comment \cite{CommentA} are also quantitatively explained by the Joule heating model.

\subsection{Overall characteristics of the nonlinear impedance unaddressed in the Comment \cite{CommentA}}

\begin{table*}[t]
\caption{Comparison between the time-varying Joule heating model \cite{FurutaPRBA} and the emergent induction scenario \cite{CommentA} regarding key overall behaviors of the observed $\Delta Z$.}
\centering
\begin{tabular}{ccc}
\hline
\begin{tabular}{c}
Key overall behavior of\\
the observed $\Delta Z$
\end{tabular}
&
\begin{tabular}{c}
Time-varying Joule\\heating model \cite{FurutaPRBA}
\end{tabular}
&
\begin{tabular}{c}
Emergent induction scenario\\discussed by Yokouchi \textit{et al}.~\cite{CommentA}
\end{tabular} \\
\hline \hline
$|\Re \Delta Z| \gg |\Im \Delta Z |$ at low $\omega$ & Expected & Not explained\\
& & \\
\begin{tabular}{c}
Similarities in the magnitudes of $\Delta Z$ and\\onset $j$ between the noncollinear\\ and ferromagnetic phases
\end{tabular}
&
Expected
& Not explained\\
& & \\
\begin{tabular}{c}
Order of magnitude of $\Delta Z$ in\\
noncollinear/ferro/paramagnetic phases
\end{tabular}
&
\begin{tabular}{c}
Predictable and\\consistent with expt.
\end{tabular}
& Not a priori explained\\
& & \\
\begin{tabular}{c}
Order of magnitude of $\omega_{\rm c}$ in\\
noncollinear/ferro/paramagnetic phases
\end{tabular}
&
\begin{tabular}{c}
Predictable and\\consistent with expt.
\end{tabular}
& Not a priori explained\\
\hline
\end{tabular}
\end{table*}

\begin{figure}
\includegraphics{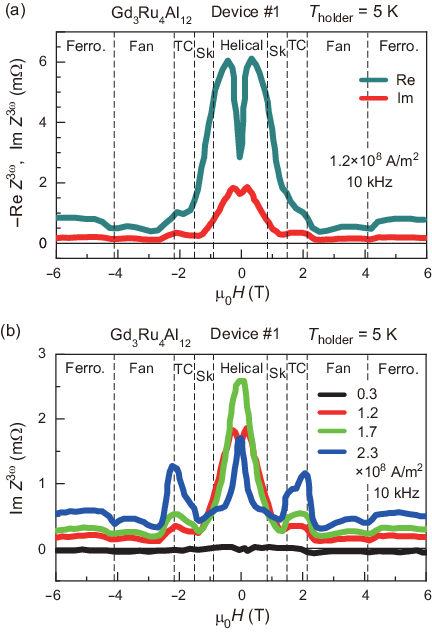}
\caption{\label{j-dep} Magnetic field dependence of (a) $-\Re Z^{3\omega}$ and $\Im Z^{3\omega}$ at 1.2$\times$10$^8$ A/m$^{2}$ and (b) $\Im Z^{3\omega}$ at various current densities for the microfabricated Gd$_{3}$Ru$_{4}$Al$_{12}$ at a sample holder temperature of 5 K. The data are reproduced from Fig.~3 in the literature \cite{YokouchiArxivA}. }
\end{figure}

In addition to the most serious problem of the negative sign of inductance \cite{YokouchiNatureA}, the overall behaviors of the observed nonlinear impedance remain unexplained in terms of the emergent induction scenario, as summarized in Table 1.

\textit{Item I}. Overall, $|\Re \Delta Z| \gg |\Im \Delta Z|$ holds at low $\omega$ [see Fig.~\ref{j-dep}(a), although 10 kHz is close to the cutoff frequency of this microfabricated specimen of Gd$_{3}$Ru$_{4}$Al$_{12}$, $\approx$30 kHz]. Yokouchi \textit{et al}.~did not explain why $\Re \Delta Z$ should be much greater than $\Im \Delta Z$ in terms of the emergent induction scenario \cite{CommentA}. This relationship is naturally expected if time-varying Joule heating is relevant to the observed $\Delta Z$ \cite{FurutaPRBA}.

\begin{figure}
\includegraphics{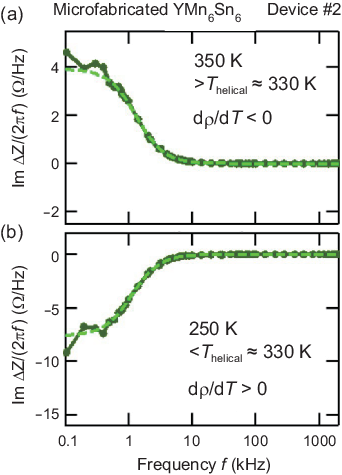}
\caption{\label{Frequency} Frequency dependence of the imaginary part of nonlinear impedance for the YMn$_6$Sn$_6$ microfabricated device at (a) 350 K ($> T_{\rm helical} \approx 330$ K) and (b) 250 K ($< T_{\rm helical} \approx 330$ K). The sign of ${\rm d}R/{\rm d}T$ is negative at 350 K and positive at 250 K \cite{FurutaPRBA, KitaoriPNASA}. The broken lines are the results of the fitting using a Debye-like frequency dispersion. The measurements were performed with $j = 2.5\times 10^8$ A/m$^2$ along the $c$-axis under zero magnetic field.
The panels are reproduced from Fig.~3.16 in Ref.~\cite{KitaoriDoctorA}.}
\end{figure}

\textit{Item II}. The magnitudes of $\Im \Delta Z$ (and $\Re \Delta Z$) for the ferromagnetic phase and the noncollinear magnetic phases, such as the helical, skyrmion, transverse conical (TC), and fan phases, are similar, as shown in Fig.~\ref{j-dep}(b). The magnitudes of the onset current density of the nonlinear impedance are also similar among all the magnetic phases. In the Comment \cite{CommentA}, they considered magnetic-phase-dependent mechanisms of $\Im \Delta Z$: depinning transition of a phason mode in the helical phase and fluctuation-induced emergent electric fields in the ferromagnetic phase. The latter mechanism refers to a theoretical study \cite{fluctuationsA}, which examines only a linear-response impedance. Yokouchi \textit{et al.}~did not discuss why the magnitudes of the nonlinear impedance and onset current density are similar between the ferro- and noncollinear magnetic phases. In contrast, if time-varying Joule heating is relevant to the observed $\Delta Z$, the similarities between the ferro and noncollinear magnetic phases are expected unless resistance values and their temperature derivatives depend greatly on magnetic order \cite{FurutaPRBA}. We also note that in YMn$_6$Sn$_6$, which is the second material reported by the same group as an ``emergent inductor'', $\Im \Delta Z$ is observed in the paramagnetic phase above the helimagnetic ordering temperature $T_{\rm helical}$. Its magnitude is similar between 350 K ($> T_{\rm helical} \approx 330$ K) and 250 K ($< T_{\rm helical}$), as shown in Fig.~\ref{Frequency} \cite{KitaoriPNASA, KitaoriDoctorA}, whereas the sign of $\Im \Delta Z$ is reversed. The time-varying Joule heating model states that $\Im \Delta Z > 0$ if ${\rm d}R/{\rm d}T < 0$ and that $\Im \Delta Z < 0$ if ${\rm d}R/{\rm d}T > 0$ \cite{FurutaPRBA}. Given negative and positive ${\rm d}R/{\rm d}T$ at 350 and 250 K, respectively \cite{FurutaPRBA, KitaoriPNASA}, the observed sign reversal is reasonably explained by the Joule heating model. Additionally, in FeSn$_2$, which was claimed by Yokouchi and coworkers to exhibit ``emergent induction,'' $\Im \Delta Z$ is observed in both the paramagnetic and collinear antiferromagnetic phases \cite{FeSn2A}. These observations highlight the tendency for a magnetic order to be irrelevant for the observed $\Im \Delta Z$.

\textit{Item III}. Yokouchi \textit{et al.}~did not verify whether the order of magnitude of the observed $\Delta Z$ can be explained by the emergent induction scenario \cite{YokouchiNatureA, CommentA}. The time-varying Joule heating model can predict the order of magnitude of $\Delta Z$ \cite{FurutaPRBA}. The estimate agrees with the experimental observations \cite{FurutaPRBA}. A good agreement is also seen for new data presented in the Comment \cite{CommentA}, which is discussed later.

\textit{Item IV}. The $\omega$--dependence of $\Delta Z$ in noncollinear magnetic phases has a cutoff frequency, $\omega_{\rm c}/(2\pi)$, as low as $\sim$10 kHz for the micrometer-sized fabricated Gd$_{3}$Ru$_{4}$Al$_{12}$ \cite{YokouchiNatureA} and $\sim$1 kHz for the micrometer-sized fabricated YMn$_{6}$Sn$_{6}$ \cite{KitaoriPNASA, KitaoriDoctorA}. In addition, in YMn$_6$Sn$_6$, the cutoff frequency $\sim$1 kHz is common to the para- and helimagnetic phases [Figs.~\ref{Frequency}(a)(b)] \cite{KitaoriDoctorA}. Yokouchi \textit{et al.}~merely ascribed the low cutoff frequencies observed in the noncollinear magnetic phases of Gd$_{3}$Ru$_{4}$Al$_{12}$ and YMn$_6$Sn$_6$ to the pinning of phason modes, without discussing whether these cutoff frequencies are reasonable values or why the cutoff frequency in YMn$_6$Sn$_6$ is common to the para- and helimagnetic phases. To the best of our knowledge, characteristic spin dynamics of 1 kHz have never been reported for the paramagnetic phase of magnetic materials, implying that the observed $\Im \Delta Z$ cannot be explained in terms of spin dynamics/fluctuations. In contrast, we experimentally demonstrated that the cutoff frequency of the Joule-heating-induced nonlinear impedance can be as low as $\sim$10 kHz for microfabricated samples and $\sim$1 Hz for bulk samples \cite{FurutaPRBA}. In other words, the time-varying Joule heating model explains why low cutoff frequencies are commonly observed in $\Delta Z$ measurements regardless of the magnetism of the conductor \cite{YokouchiNatureA, KitaoriPNASA, FeSn2A, FurutaPRBA}.

\textit{Summary}. The time-varying Joule heating model provides perspective and overall explanations as to why the experiments show similar magnitude, onset current density, and cutoff frequency of the nonlinear impedance between the noncollinear, ferro, and paramagnetic phases. Furthermore, the magnitudes of $\Delta Z$ and $\omega_{\rm c}/(2\pi)$ are predictable by the Joule heating model and are found to be in agreement with the experimental results \cite{FurutaPRBA}. At present, the emergent induction scenario does not provide a picture that explains the overall characteristics listed in Table 1, and this shortcoming makes the discussion by Yokouchi \textit{et al.}~speculative.

\subsection{Comments on the arguments in the Comment \cite{CommentA}}

\begin{figure}
\includegraphics{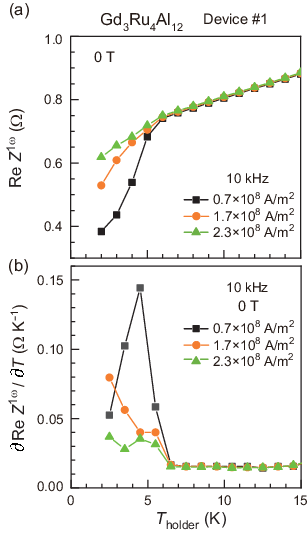}
\caption{\label{Resistivity} Temperature dependence of (a) $\Re Z^{1\omega}$ and (b)
$\partial \Re Z^{1\omega}/\partial T$ for the Gd$_3$Ru$_4$Al$_{12}$ microfabricated device used in Ref.~\cite{YokouchiNatureA}.}
\end{figure}

In the Comment \cite{CommentA}, Yokouchi \textit{et al}.~argue that the observed $\Delta Z$ behavior is inconsistent with the prediction of the time-varying Joule heating model (for instance, $\Im Z^{3\omega} \propto -\Re Z^{3\omega} \propto {\rm d}R/{\rm d}T$). However, we note that in Ref.~\cite{FurutaPRBA}, we derived the equations for the Joule-heating-induced impedance by considering only the lowest-order contribution of Joule heating. If ${\rm d}R/{\rm d}T$ appreciably changes as the applied current density increases, the data should be analyzed using a more elaborate model that involves the temperature dependences of ${\rm d}R/{\rm d}T$ and specific heat, temperature inhomogeneity in the sample, and so on. In fact, as shown in Fig.~\ref{Resistivity}, the Gd$_{3}$Ru$_{4}$Al$_{12}$ specimen under consideration has ${\rm d}R/{\rm d}T$ that is sensitive to the applied current density in the range between 0.7 and 1.7$\times$10$^{8}$ A/m$^{2}$ at 5 K and 0 T, indicating that the data should not be analyzed within the scope of the lowest-order Joule-heating model.

\begin{figure}
\includegraphics{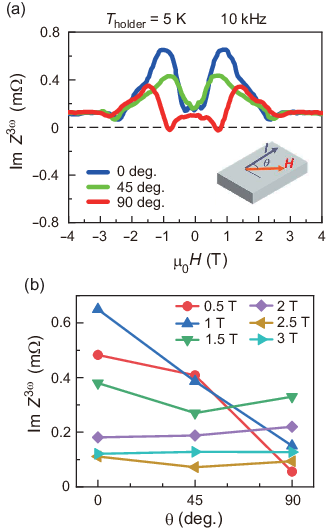}
\caption{\label{Angle} (a) Magnetic field dependence of $\Im Z^{3\omega}$ at 1.2$\times$10$^8$ A/m$^{2}$ at selected in-plane field angles. The data are reproduced from Fig.~1 in the Comment \cite{CommentA}. (b) In-plane field angle dependence of $\Im Z^{3\omega}$ at selected magnetic fields. The data in Panel (b) are taken from Panel (a).}
\end{figure}

In Fig.~1 of the Comment \cite{CommentA}, referring to the data at 5 K, they argue that the relationship of $\Im Z^{3\omega} \propto -\Re Z^{3\omega} \propto \partial R/\partial T$ is not well satisfied; therefore, this observation should indicate an emergent induction mechanism. As mentioned above, such a complicated $\partial R/\partial T$-behavior requires a more elaborate Joule heating model. We also note that a similar deviation from the minimal model can be observed around a phase transition in the paramagnetic metal MoTe$_{2}$ \cite{FurutaPRBA, FurutaArxivA}.

This deviation from the lowest-order Joule heating model does not necessarily mean that the spin-transfer-torque (STT) mechanism is relevant to the observed $\Delta Z$. Figure 1 of the Comment \cite{CommentA} displays the in-plane magnetic field dependence of $\Im Z^{3\omega}$ at selected angles, $\theta$, with respect to the current direction, which is reproduced in Fig.~\ref{Angle}(a). Yokouchi \textit{et al}.~argue that $\Im Z^{3\omega}$ at approximately 1 T decreases as $\theta$ varies from $0^{\circ}$ to $90^{\circ}$, and this trend is consistent with what is expected from the STT-induced emergent induction. However, if one plots the $\theta$ dependence of $\Im Z^{3\omega}$ at various magnetic fields [Fig.~\ref{Angle}(b)], one can find that such systematic behavior emerges only at some limited magnetic fields. For instance, at 1.5 T, $\Im Z^{3\omega}(45^{\circ}) < \Im Z^{3\omega}(90^{\circ})$. The overall behavior contradicts the relationship derived from the STT-induced emergent induction, namely $\Im Z^{3\omega}(0^{\circ}) > \Im Z^{3\omega}(45^{\circ}) > \Im Z^{3\omega}(90^{\circ})$.

Notably, although the sign of $\Re Z^{3\omega}$ is negative overall, positive $\Re Z^{3\omega}$ is exceptionally observed around the multi-to-monodomain transition field of the helimagnetic order under an in-plane field at $\theta = 90^{\circ}$. Yokouchi \textit{et al}.~argue that this behavior cannot be explained by Joule heating and should originate from a mechanism that is unique to the multi-to-monodomain transition point. This statement appears to be correct, and we agree that as long as $\partial R/\partial T > 0$, it is difficult to explain the positive $\Re Z^{3\omega}$ in terms of Joule heating. Therefore, this unique behavior deserves further investigation. Yokouchi \textit{et al}.~also state that ``a more comprehensive understanding of the origins of $\Re Z^{3\omega}$ is beyond the scope of this paper and remains a task for future research.'' In any case, the positive $\Re Z^{3\omega}$ is rather exceptional in the overall dataset. The peculiar behavior observed at an in-plane field of $\approx$1 T with $\theta = 90^{\circ}$ should be considered separately.

In Fig.~2 of the Comment \cite{CommentA}, they referred to the lowest-order Joule-heating model and estimated a possible temperature increase, $\Delta T$, at 5 K and 0 T under 1.3$\times$10$^{8}$ A/m$^{2}$ to be nearly zero. We note that the estimate of $\Delta T$ by Yokouchi \textit{et al}.~lacks a valid basis because the lowest-order Joule-heating model cannot be applied to the data analysis at 5 K and 0 T. A more reliable, model-free estimate of $\Delta T$ can be obtained by referring to the shift in the magnetic transition field under the application of current. In Ref.~\cite{YokouchiArxivA}, Yokouchi \textit{et al}.~present the magnetic field dependence of $\Re Z^{1\omega}$, showing that the helical-to-conical transition field detected under 1.2$\times$10$^{8}$ A/m$^{2}$ is different from that under 0.7$\times$10$^{8}$ A/m$^{2}$ (for more details, see Fig.~8 of Ref.~\cite{FurutaArxivA}). This observation indicates that even at 1.2$\times$10$^{8}$ A/m$^{2}$, the temperature increase is not negligible, and $\Delta T$ amounts to $\approx$0.09 K for the measurement at 5 K. Referring to the $\partial R/\partial T $ value of 15 m$\Omega/\rm{K}$ at 7--15 K [Fig.~\ref{Resistivity}(b)], where the application of the lowest-order Joule heating model appears valid because $\partial R/\partial T$ does not vary significantly with the current, we obtain 2.25 m$\Omega$ as an estimate of the Joule-heating-induced $\Re \Delta Z^{1\omega}$ at low $\omega$. This value is in good agreement with the new data reported in the Comment, 2--4 m$\Omega$ at 8--18 K (see Fig.~2 of the Comment \cite{CommentA}). Thus, their new data are quantitatively explained from the perspective of Joule heating. We see no discussion of how the observed $\Delta Z$ value is justified in terms of the emergent induction in the Comment \cite{CommentA}.


\newpage

\begin{center}
\textbf{\large Reply to arXiv:2407.15682, submitted by Yokouchi \textit{et al.}}\\[1.5em]

Soju Furuta,$^{1}$ Wataru Koshibae,$^{2}$ Taka-hisa Arima,$^{2,3}$ and Fumitaka Kagawa$^{1,2,*}$\\[1em]

\textit{$^{1}$Dept.~of Physics, Institute of Science Tokyo, Tokyo 152-8551, Japan}\\
\textit{$^{2}$RIKEN Center for Emergent Matter Science (CEMS), Wako 351-0198, Japan}\\
\textit{$^{3}$Dept.~of Advanced Materials Science, University of Tokyo, Kashiwa 277-8561, Japan}\\[1em]

(Dated: October 5, 2024)
\end{center}


\setcounter{page}{1}

\begin{center}
\begin{minipage}{1.0\linewidth}
\noindent
\begin{center}
    \large Abstract
\end{center}
We respond to the ``Comment on ``Reconsidering the nonlinear emergent inductance: time-varying Joule heating and its impact on the AC electrical response'''', submitted by Yokouchi \textit{et al.}~(arXiv:2407.15682), which refers to our preprint by Furuta \textit{et al} (arXiv:2407.00309). In our preprint, we argued that the nonlinear impedance reported for materials with noncollinear magnetic textures, Gd$_3$Ru$_4$Al$_{12}$, YMn$_6$Sn$_6$, and FeSn$_2$, can be attributed to the considerable contribution of Joule-heating-induced AC electrical response, rather than the emergent electric field (EEF) due to the current-induced-dynamics of magnetic textures. In the comment by Yokouchi \textit{et al.}, the authors presented new data and concluded that the nonlinear impedance previously reported for Gd$_3$Ru$_4$Al$_{12}$ and YMn$_6$Sn$_6$ was not due to Joule heating but to the EEF. After reviewing their data and arguments, we found that (i) they misunderstood the applicability of the Joule heating model discussed in our preprint (arXiv:2407.00309), (ii) the data they presented in the comment are affected by extrinsic superconductivity caused during the microfabrication of specimens using focused ion beam (FIB) methods, and (iii) they misinterpreted their new data regarding the magnetic-field-angle dependence of the nonlinear impedance. Thus, we maintain that the nonlinear impedance reported in the literature and the comment includes a considerable impact from Joule heating. Furthermore, we demonstrate that for Gd$_3$Ru$_4$Al$_{12}$, the cross-sectional area dependence of the nonlinear impedance refutes the EEF scenario, and that the sign change of nonlinear resistance observed by Yokouchi \textit{et al.}~is not allowed from the perspective of stability associated with the current-driven EEF.
\end{minipage}
\end{center}
\vspace{2em}


\setcounter{figure}{0}
\setcounter{table}{0}

\subsection{I. Introduction}

An emergent inductor in a material with noncollinear magnetic textures, which was theoretically proposed by Nagaosa \cite{NagaosaJJAP}, is a new class of inductors. The mechanism is a combination of the spin-transfer torque (STT) effect \cite{STT1, STT2} and the resulting emergent electric field (EEF) \cite{Volovik, BarnesPRL2007}. When an AC electric current below the depinning threshold is applied to the magnetic texture, oscillatory elastic deformation occurs; thus, the time evolution of the magnetic texture results in an oscillatory EEF response that is out-of-phase with respect to the applied AC current. After the theoretical proposal \cite{NagaosaJJAP}, experimental studies on Gd$_3$Ru$_4$Al$_{12}$ \cite{YokouchiNature}, YMn$_6$Sn$_6$ \cite{KitaoriPNAS}, and FeSn$_2$ \cite{YokouchiArxiv} were reported, and it was argued that nonlinear imaginary impedance under a relatively large current density ($\sim$$10^{8}$ A~m$^{-2}$) in the presence of noncollinear magnetic textures originates from the nonlinear response of the spin system. However, in our preprint \cite{PrepriFuruta}, we reached the following two conclusions regarding the experimental observations in the literature \cite{YokouchiNature, KitaoriPNAS, YokouchiArxiv}.

\textit{Conclusion I}. The nonlinear imaginary impedance reported previously is accompanied by a much larger nonlinear real impedance; in other words, the nonlinearly induced impedance has resistor-like (i.e., dissipative) characteristics rather than nondissipative inductor-like characteristics. This conclusion is in contrast with previous claims \cite{YokouchiNature, KitaoriPNAS, YokouchiArxiv} that the observed nonlinear impedance has an inductor-like response. 

\textit{Conclusion II}. Joule heating plays a considerable role in the resistor-like nonlinear impedance. 

During the review process of our preprint \cite{PrepriFuruta} by \textit{Physical Review B}, Yokouchi \textit{et al}.~from the Tokura group commented on our preprint and refuted the conclusion \textit{II} \cite{Condmat}. This paper is our response to their comment \cite{Condmat}. 

This paper is organized as follows: In Section~II, we discuss the two different models that were introduced in the first experimental report \cite{YokouchiNature} and the comment \cite{Condmat} to explain the observed nonlinear impedance and show that these models cannot explain the experiments \cite{YokouchiNature, KitaoriPNAS, YokouchiArxiv, Condmat}. In Section~III, we explain why the dissipative nonlinear impedance observed in the previous experiments \cite{YokouchiNature, KitaoriPNAS, YokouchiArxiv} was misinterpreted as nondissipative nonlinear impedance. In Sections~IV--VI, we discuss from various perspectives whether the dissipative nonlinear impedance is due to Joule heating or the EEF. In Section~IV, we present the cross-sectional area dependence of the nonlinear imaginary impedance and argue that the data refute the EEF-based interpretation. In Section~V, we discuss the applicability of the minimal Joule heating model introduced in our preprint \cite{PrepriFuruta} because we find misleading arguments in the comment \cite{Condmat}. In Section~VI, we comment on each figure presented in the comment \cite{Condmat}. In particular, we point out problems in the data interpretation and data handling and show that the new data are also consistent with the Joule-heating-based interpretation. In Section~VII, we summarize this document.

\subsection{II. Consideration for the emergent-electric-field-induced nonlinear impedance from a macroscopic perspective}

In general, there is a close relationship between the mechanism and signs of impedance. This relationship can be discussed in terms of the stability requirements of the electrical response. For instance, the stabilities of a resistor and an inductor are guaranteed by positive values of the linear-response resistance and inductance, respectively. Given that the EEF response is expected to be independent of the linear-response resistance \cite{NagaosaJJAP, Volovik, BarnesPRL2007}, the EEF response must satisfy the conditions that guarantee a stable electrical response. In the literature \cite{YokouchiNature, KitaoriPNAS, YokouchiArxiv} and comment \cite{Condmat}, the authors claimed that the linear-response EEF is negligibly small, and the nonlinearly induced EEF dominates the observed electrical response. Therefore, if their claim is correct, then the dominant nonlinear EEF signal must meet the stability conditions, implying that the signs of the Fourier components of the nonlinearly induced electrical response, such as $\Re \rho^{1\omega}$, $\Re \rho^{3\omega}$ (whose definitions are given below), $\Im \rho^{1\omega}$ and $\Im \rho^{3\omega}$, have some constraints. In this section, we discuss this issue by exploring the case in which the EEF originates from a nonlinear inductive mechanism with no threshold value \cite{YokouchiNature} and the case in which the EEF originates from a nonlinear resistive mechanism with a threshold current density \cite{Condmat}. We abbreviate these cases as the ``nonlinear inductive EEF model'' and ``nonlinear resistive EEF model,'' respectively. The details of these two models are explained below.

\subsection{A. Nonlinear inductive EEF model with no threshold value}

In this subsection, we discuss the sign of the impedance derived from the phenomenological nonlinear inductor model introduced in Ref.~\cite{YokouchiNature}. In Ref.~\cite{YokouchiNature}, the current-induced deformation of pinned magnetic textures under AC current density, $j(t) = j_0 \sin \omega t$, is considered, and the time evolution of the EEF, $E_{\rm e}(t)$, is assumed to be described by the following equation:
\begin{align}
\label{EEF}
E_{\rm e}(t) = \Big( \tilde{L}_0 + \tilde{L}_2 j(t)^2 + \tilde{L}_4 j(t)^4 + \cdots \Big) \frac{{\rm d}j}{{\rm d}t}, 
\end{align}
where $\tilde{L}_i$ $(i = 0, 2, 4, \cdots)$ are constants. We consider the characteristics associated with Eq.~(\ref{EEF}) from a macroscopic perspective. Equation (\ref{EEF}) does not include an energy-dissipation (i.e., resistive) term. Therefore, energy conservation requires that during the AC cycle, the work done by the external power supply against the inductive counter-electromotive force due to the EEF should be stored in the magnetic texture and then returned to the external power supply \cite{Furuta1, Furuta2}. As long as the elastic deformation of the magnetic texture can sufficiently follow the time varying current, the current-induced energy density stored in the magnetic texture, $u(t)$, is given by:
\begin{align}
\label{Energy}
u(t) &\approx \int_0^{t}{\rm d}t'j(t')E_{\rm e}(t') \nonumber \\
&= \frac{1}{2}\tilde{L}_0 j(t)^2 + \frac{1}{4}\tilde{L}_2 j(t)^4 + \frac{1}{6}\tilde{L}_4 j(t)^6 + \cdots.
\end{align}
For the system to remain stable, $u(j)$ must exhibit a local minimum only at $j=0$ in the considered range of $j$. If another local minimum exists at a finite $j$, the system includes an unstable solution (see the Supplemental Material in Ref.~\cite{PrepriFuruta}). Therefore, ${\rm d}u/{\rm d}j$ should be positive (negative) for arbitrary positive (negative) $j$; otherwise, the energy would decrease as the $|j|$ increases, resulting in a spontaneous $|j|$ increase. Thus, 
\begin{align}
\frac{{\rm d}u}{{\rm d}j} = j(\tilde{L}_0 + \tilde{L}_2 j^2 + \tilde{L}_4 j^4 + \cdots) &> 0 \hspace{0.5cm} \rm if \hspace{0.2cm} \textit j > 0. \nonumber \\
\label{Stability}
\therefore \tilde{L}_0 + \tilde{L}_2 j^2 + \tilde{L}_4 j^4 + \cdots &> 0.
\end{align}
Equation (\ref{Stability}) demonstrates that for the system to remain stable (i.e., not self-oscillatory), the coefficient of ${\rm d}j/{\rm d}t$ in Eq.~(\ref{EEF}) should be positive for any instant (see also the Supplemental Material in Ref.~\cite{PrepriFuruta}).

Below, for simplicity, we consider nonlinearity in Eq.~(\ref{EEF}) up to the $\tilde{L}_4 j(t)^4$ term. The stability condition for this case is given as $\tilde{L}_0 > 0$, $\tilde{L}_4 > 0$, and $\tilde{L}_2 > -2(\tilde{L}_0\tilde{L}_4)^{1/2}$. As long as the system remains stable, the EEF response under $j(t) = j_0 \sin \omega t$ is thus given by:
\begin{align}
\label{e-field_response}
\frac{E_{\rm e}(t)}{j_0} = \omega \times \Bigg[ &\tilde{L}_0 \cos \omega t \nonumber \\
&+ \tilde{L}_2 j_0^2 \left( \frac{1}{4}\cos \omega t - \frac{1}{4}\cos 3\omega t \right) \nonumber \\
&+ \tilde{L}_4 j_0^4 \left( \frac{1}{8}\cos \omega t - \frac{3}{16}\cos 3\omega t + \frac{1}{16}\cos 5\omega t \right) \Bigg].
\end{align}
Thus, the EEF-derived electrical responses at the 1$\omega$, 3$\omega$ and 5$\omega$ components are given as follows:
\begin{align}
\Im \rho_{\rm e}^{1\omega} &\equiv \frac{\Im E_{\rm e}^{1\omega}}{j_0} = \omega \left( \tilde{L}_0 + \frac{1}{4}\tilde{L}_2 j_0^2 + \frac{1}{8}\tilde{L}_4 j_0^4 \right), \\
\Im \rho_{\rm e}^{3\omega} &\equiv \frac{\Im E_{\rm e}^{3\omega}}{j_0} = \omega \left( -\frac{1}{4}\tilde{L}_2 j_0^2 -\frac{3}{16}\tilde{L}_4 j_0^4 \right), \\
\Im \rho_{\rm e}^{5\omega} &\equiv \frac{\Im E_{\rm e}^{5\omega}}{j_0} = \omega \left( \frac{1}{16}\tilde{L}_4 j_0^4 \right).
\end{align}
The system stability conditions $\big[$i.e., $\tilde{L}_0 > 0$, $\tilde{L}_4 > 0$, and $\tilde{L}_2 > -2(\tilde{L}_0\tilde{L}_4)^{1/2} \big]$ guarantee that $\Im \rho_{\rm e}^{1\omega}$ is positive for arbitrary $j_0$.

Equation (\ref{e-field_response}) describes the EEF response for the case where the current-induced elastic deformation of the magnetic texture exactly follows the instantaneous current value in the low-frequency limit. At a nonzero frequency, the magnetic texture deformation may be accompanied by a phase delay relative to the instantaneous driving force (electric current). To account for this effect, we introduce positive values, $\alpha_1(\omega) > 0, \alpha_3(\omega) > 0$, and $\alpha_5(\omega) > 0$, representing the phase delays for the 1$\omega$, 3$\omega$, and 5$\omega$ responses, respectively. Then, as long as the low-frequency regime with small phase delays $\big[$i.e., $\alpha_n(\omega) \ll 1$ ($n = 1,3,5$)$\big]$ is considered, $\cos n\omega t$  in Eq.~(\ref{e-field_response}) are replaced by $\cos \big( n\omega t - \alpha_n (\omega) \big) \approx \cos n\omega t + \alpha_n (\omega) \sin n\omega t$. Thus, the real parts of the EEF-derived electrical responses are given as:
\begin{align}
\Re \rho_{\rm e}^{1\omega} &\equiv \frac{\Re E_{\rm e}^{1\omega}}{j_0} \approx \alpha_1(\omega) \Im \rho_{\rm e}^{1\omega} = \alpha_1(\omega) \omega \left( \tilde{L}_0 + \frac{1}{4} \tilde{L}_2 j_0^2 + \frac{1}{8} \tilde{L}_4 j_0^4  \right), \\
\Re \rho_{\rm e}^{3\omega} &\equiv \frac{\Re E_{\rm e}^{3\omega}}{j_0} \approx \alpha_3(\omega) \Im \rho_{\rm e}^{3\omega} = \alpha_3(\omega) \omega \left( - \frac{1}{4}\tilde{L}_2 j_0^2 - \frac{3}{16}\tilde{L}_4 j_0^4 \right), \\
\Re \rho_{\rm e}^{5\omega} &\equiv \frac{\Re E_{\rm e}^{5\omega}}{j_0} \approx \alpha_5(\omega) \Im \rho_{\rm e}^{5\omega} = \alpha_5(\omega) \omega \left( \frac{1}{16}\tilde{L}_4 j_0^4 \right).
\end{align}
Note that as long as the low-frequency regime is considered, $|\Im \rho_{\rm e}^{n\omega}| \gg |\Re \rho_{\rm e}^{n\omega}|$. This is a natural consequence of the fact that the impedance of an inductor is nondissipative. 

\begin{table*}
\caption{Signs of the nonlinear impedance from the nonlinear inductive EEF model, nonlinear resistive EEF model and moderate Joule heating model. Note that in the Joule heating model, the signs of the nonlinear impedance depend on the sign of $\frac{{\rm \partial}\rho}{{\rm \partial}T}$ (the temperature derivative of resistivity). The relationship between the real and imaginary components of the nonlinear impedance at low $\omega$ for each mechanism is also shown.}
\centering
\begin{tabular}{cccc}
\hline
& 
\begin{tabular}{c}
Nonlinear \\
\hspace{0.2cm} inductive EEF \hspace{0.2cm}
\end{tabular}
& 
\begin{tabular}{c}
Nonlinear \\
\hspace{0.2cm} resistive EEF \hspace{0.2cm}
\end{tabular}
&
\begin{tabular}{c}
Moderate \\
\hspace{0.2cm} Joule heating \cite{PrepriFuruta}
\end{tabular}\\
\hline \hline
At low $\omega$
&
\begin{tabular}{c}
\hspace{0.1cm} $| \Re \rho_{\rm e}^{1\omega} | \ll |\Im \rho_{\rm e}^{1\omega}|$ \hspace{0.1cm} \\
\hspace{0.1cm} $|\Re \rho_{\rm e}^{3\omega}| \ll |\Im \rho_{\rm e}^{3\omega}|$ \hspace{0.1cm}
\end{tabular}
&
\begin{tabular}{c}
\hspace{0.1cm} $|\Re \rho_{\rm e}^{1\omega}| \gg |\Im \rho_{\rm e}^{1\omega}|$ \hspace{0.1cm}\\
\hspace{0.1cm} $|\Re \rho_{\rm e}^{3\omega}| \gg |\Im \rho_{\rm e}^{3\omega}|$ \hspace{0.1cm}
\end{tabular}
&
\begin{tabular}{c}
\hspace{0.1cm} $|\Re \Delta \rho^{1\omega}| \gg |\Im \Delta \rho^{1\omega}|$ \hspace{0.1cm}\\
\hspace{0.1cm} $|\Re \rho^{3\omega}| \gg |\Im \rho^{3\omega}|$ \hspace{0.1cm}
\end{tabular}
\\
$\Re \rho_{\rm e}^{1\omega}$ or $\Re \Delta \rho^{1\omega}$ & $+$ & $+$ 
& 
$+$ ($\frac{{\rm \partial}\rho}{{\rm \partial}T} > 0$) / $-$ ($\frac{{\rm \partial}\rho}{{\rm \partial}T} <0$)\\
$\Im \rho_{\rm e}^{1\omega}$ or $\Im \Delta \rho^{1\omega}$ & $+$ & $-$
&
$-$ ($\frac{{\rm \partial}\rho}{{\rm \partial}T} > 0$) / $+$ ($\frac{{\rm \partial}\rho}{{\rm \partial}T} <0$)\\
$\Re \rho_{\rm e}^{3\omega}$ or $\Re \rho^{3\omega}$ & $-$ & $-$
&
$-$ ($\frac{{\rm \partial}\rho}{{\rm \partial}T} > 0$) / $+$ ($\frac{{\rm \partial}\rho}{{\rm \partial}T} <0$)\\
$\Im \rho_{\rm e}^{3\omega}$ or $\Im \rho^{3\omega}$ & $-$ & $+$
&
$+$ ($\frac{{\rm \partial}\rho}{{\rm \partial}T} > 0$) / $-$ ($\frac{{\rm \partial}\rho}{{\rm \partial}T} <0$)\\
\hline
\end{tabular}
\end{table*}

In the experiments \cite{YokouchiNature, KitaoriPNAS, YokouchiArxiv}, the imaginary part of the impedance is nearly zero in the linear-response regime. If this observation is considered on the basis of the EEF described by Eq.~(\ref{EEF}), $\tilde{L}_0 \approx 0$ and the system stability requires $\tilde{L}_2 > 0$. It follows that the signs of the Fourier components are determined as $\Im \rho_{\rm e}^{1\omega} > 0$, $\Im \rho_{\rm e}^{3\omega}< 0$, $\Re \rho_{\rm e}^{1\omega} > 0$ and $\Re \rho_{\rm e}^{3\omega} < 0$ (Table I). These signs are opposite to those observed in the previous studies \cite{YokouchiNature, KitaoriPNAS, YokouchiArxiv}, indicating that the nonlinear imaginary impedance reported in Refs.~\cite{YokouchiNature, KitaoriPNAS, YokouchiArxiv} is unrelated to the inductive behavior described by Eq.~(\ref{EEF}). In fact, as shown in our preprint \cite{PrepriFuruta}, a double-check of the raw data in Refs.~\cite{YokouchiNature, KitaoriPNAS, YokouchiArxiv} reveals that the nonlinear impedance has mainly \textit{dissipative} (i.e., resistor-like) characteristics. In Refs.~\cite{YokouchiNature, KitaoriPNAS, YokouchiArxiv}, the authors of the comment \cite{Condmat} initially interpreted the nonlinear impedance as having mainly \textit{nondissipative} characteristics and thus concluded that an emergent inductor with negative inductance was present. However, in the abstract of the comment \cite{Condmat}, the authors later stated that, ``we observe a significant real part of the nonlinear complex impedance, likely resulting from the dissipation associated with the current-driven motion of helices and domain walls''. Accordingly, the authors \cite{Condmat} appear to have revised their interpretation of the nonlinear impedance from the original reports \cite{YokouchiNature, KitaoriPNAS, YokouchiArxiv} and now agree with our conclusion \textit{I}.

\subsection{B. Nonlinear resistive model with a threshold current density}

\begin{figure*}
\includegraphics{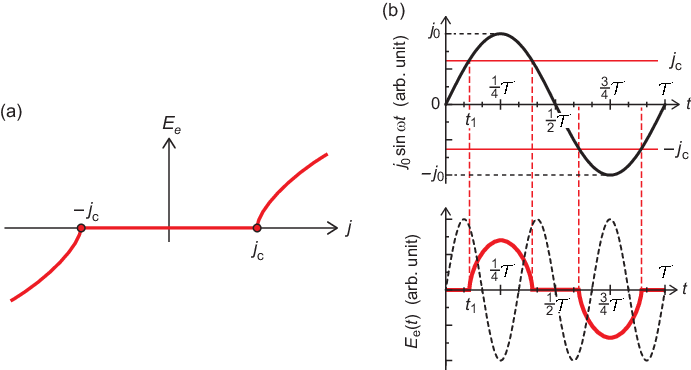}
\caption{\label{3w} (a) Schematic of the $E_{\rm e}$--$j$ characteristics expected for a depinned magnetic texture with a threshold current density $j_{\rm c}$. (b) Time profiles of $j(t)$ (upper panel) and the schematic $E_{\rm e}\big( j(t) \big)$ (lower panel). In the lower panel, the curve $\sin 3\omega t$ (dotted line) is shown as a reference for the calculation of $\Re E_{\rm e}^{3\omega}$.
}
\end{figure*}

In this subsection, we discuss the signs of the nonlinear impedance in the low-frequency regime, for the case where the EEF of dissipative characteristics originates from the current-induced depinned motion of a magnetic texture, as stated in the comment \cite{Condmat}. To this end, the $j$--$E_{\rm e}$ curve with respect to a DC current density should first be defined. A canonical $j$--$E_{\rm e}$ curve, expected for the pinning--depinning transition of a magnetic texture \cite{Depinning}, is shown in Fig.~\ref{3w}(a). Below the threshold current density $j_{\rm c}$, the magnetic texture remains pinned, and thus, no time evolution is observed in the steady state, leading to $E_{\rm e} = 0$. For $j > j_{\rm c}$, it exhibits a steady flow of the depinned magnetic texture and thus finite $E_{\rm e}$ appears, as depicted by the $j$--$E_{\rm e}$ curve. It should be noted that $E_{\rm e}$ must remain positive for $j > j_{\rm c}$, as depinned motion must be accompanied by an energy dissipation. Furthermore, the $j$--$E_{\rm e}$ curve satisfies the general stability condition of a $j$--$E$ curve (i.e., nonnegative values of ${\rm d} E/{\rm d}j$ for arbitrary $j$ \cite{Stability}; i.e, the $j$--$E_{\rm e}$ curve monotonically increases for $j > j_{\rm c}$). Here, the so-called Walker breakdown, which occurs in highly nonlinear regimes \cite{Walker}, is not considered, as it is accompanied by self-oscillatory response, which is out of scope of the paper. 

Having established the $j$--$E_{\rm e}$ curve with respect to a DC current density, we now consider the electrical response under the AC current density, $j(t) = j_0 \sin \omega t$ with $j_0 > j_{\rm c}$. The time profiles of $j(t)$ and the resulting $E_{\rm e}\big( j(t) \big)$ are schematically illustrated in Fig.~\ref{3w}(b). $\Re E_{\rm e}^{1\omega} \big[ \equiv \frac{2}{\mathcal{T}}\int^\mathcal{T}_0 dt \; E_{\rm e}(j_0 \sin \omega t) \sin \omega t$, where $\mathcal{T} \equiv 2\pi/\omega \big]$ is obviously positive because $E_{\rm e}(t) \sin \omega t \geq 0$ holds at any given moment. Given that the $j$--$E_{\rm e}$ curve monotonically increases for $j > j_{\rm c}$, the following inequality holds for $\Re E_{\rm e}^{3\omega}$:
\begin{align}
\Re E_{\rm e}^{3\omega} &\equiv \frac{2}{\mathcal{T}}\int^\mathcal{T}_0 dt \; E_{\rm e}(j_0 \sin \omega t) \sin 3\omega t \nonumber \\
&=4\times \frac{2}{\mathcal{T}} \left[ \int^{\mathcal{T}/6}_{t_1} dt + \int^{\mathcal{T}/4}_{\mathcal{T}/6} dt \right] E_{\rm e}(j_0 \sin \omega t) \sin 3\omega t \nonumber \\
\label{Square}
&\leq \frac{8}{\mathcal{T}} E_{\rm e} \left( j_0 \sin \omega \frac{\mathcal{T}}{6} \right) \left[ \int^{\mathcal{T}/6}_{t_1} dt + \int^{\mathcal{T}/4}_{\mathcal{T}/6} dt \right] \sin 3\omega t \\
\label{Negative}
&= \frac{8}{\mathcal{T}} E_{\rm e}\left( \frac{\sqrt{3}}{2}j_0 \right) \frac{\cos 3\omega t_1}{3\omega}.
\end{align}
The definition of $t_1$ is illustrated in Fig.~\ref{3w}(b). For $t_1$ values satisfying $\frac{1}{12}\mathcal{T} < t_1 < \frac{1}{4}\mathcal{T}$, Eq.~(\ref{Negative}) is always negative. This condition for $t_1$ corresponds to $j_0 < 2 j_c$. Thus, for $j_c < j_0 < 2 j_c$, $\Re E_{\rm e}^{3\omega}$ should invariably be negative. Note that $j_0 \geq 2 j_c$ does not necessarily lead to a positive $\Re E_{\rm e}^{3\omega}$, as is obvious in Eqs.~(\ref{Square}) and (\ref{Negative}).

For the imaginary components of the current-induced resistor-like EEF responses in the low-frequency regime, the positive values of the phase delay, $\alpha_n(\omega) > 0$ (see Section~II A), should be considered:
\begin{align}
E_{\rm e}(t)/j_0 &\approx \sum (\Re \rho_{\rm e}^{n\omega}) \sin \big( n\omega t - \alpha_n (\omega) \big) \hspace{1cm} (n = 1,3, \cdots) \nonumber \\
&\approx \sum (\Re \rho_{\rm e}^{n\omega}) \big[\sin n\omega t - \alpha_n (\omega) \cos n\omega t \big]. \nonumber \\
\therefore \Im \rho_{\rm e}^{n\omega} &\approx -\alpha_n (\omega) \Re \rho_{\rm e}^{n\omega},
\end{align}
Note that as long as the low-frequency regime is considered, $|\Re \rho_{\rm e}^{n\omega}| \gg |\Im \rho_{\rm e}^{n\omega}|$. This is a natural consequence of the fact that impedance of a resistor is dissipative. Thus, for the nonlinear resistive EEF caused by the current-induced depinned motion of the magnetic texture, $\Re \rho_{\rm e}^{1\omega} > 0$, $\Re \rho_{\rm e}^{3\omega}< 0$, $\Im \rho_{\rm e}^{1\omega} < 0$, and $\Im \rho_{\rm e}^{3\omega} > 0$ hold (Table I) unless a highly depinned regime is considered. As discussed in Section~VI~C, some of the data in Ref.~\cite{Condmat} exhibit $\Re \rho_{\rm e}^{3\omega}> 0$, which is not consistent with the resistive EEF model with a threshold current density.

The signs derived from the nonlinear resistive EEF model are the same as those expected for the nonlinear AC electrical response of a material with ${\rm \partial}\rho/{\rm \partial}T > 0$ ($T$ represents the temperature) under moderate Joule heating (Table I). Here, ``moderate'' means that Joule heating is of a sufficiently small magnitude to allow the model discussed in Ref.~\cite{PrepriFuruta} to be applied, and $\Delta \rho^{1\omega}$ in Table I represents a change from the linear-response complex resistivity of a material, $\rho^{1\omega}$. Note that for the case of Joule-heating-induced nonlinear AC electrical response, opposite signs are observed for a material with ${\rm \partial}\rho/{\rm \partial}T < 0$ (Table I). The signs of $\Delta \rho^{1\omega}$ and $\rho^{3\omega}$ under moderate Joule heating are simply determined by whether the value of $E/j$ becomes larger or smaller at higher current densities. The sign of the Joule-heating-induced $\Delta \rho^{1\omega}$ is allowed to be either positive or negative because the stability of the system is guaranteed by a positive value of the linear-response resistivity.

\subsection{III. Nonlinear impedance seen from the complex plane}

\begin{figure*}
\includegraphics{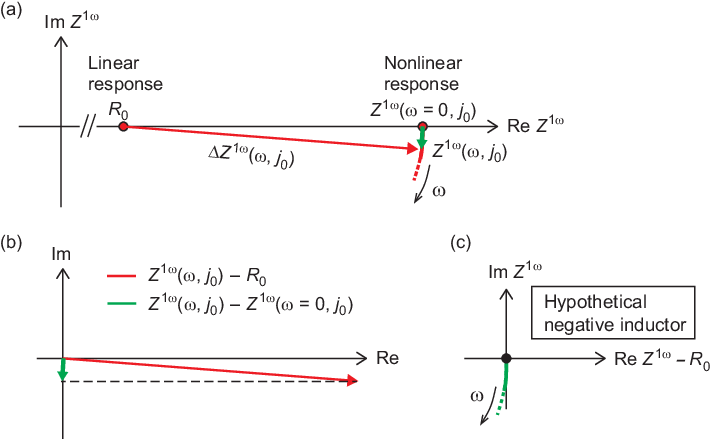}
\caption{\label{ImpedancePlane} (a) Complex plane representation of the linear and nonlinear impedances. (b) The starting points of the two arrows (red and green) shown in (a) are moved to the origin. (c) Behavior expected for a hypothetical negative inductor. Note, however, that a negative inductor is unstable, as discussed in the Supplemental Material of Ref.~\cite{PrepriFuruta}. 
}
\end{figure*}

We explain why the dissipative nonlinear impedance $Z^{1\omega}$ was misinterpreted as nondissipative in Refs.~\cite{YokouchiNature, KitaoriPNAS, YokouchiArxiv}. For this purpose, it is helpful to refer to the complex plane and consider the linear and nonlinear impedances as complex vectors, as shown in Fig.~{\ref{ImpedancePlane}}(a) and (b). In the experiments \cite{YokouchiNature, KitaoriPNAS, YokouchiArxiv}, the authors found that $\Im Z^{1\omega}$ is negligibly small in the linear response for frequencies of $\omega/(2\pi)$ less than 1 MHz, meaning that the linear-response impedance $Z^{1\omega}(\omega)$ (i.e., independent of the current density, $j_0$) can be approximated by an $\omega$-independent real value, $R_0$ (i.e., resistance in the usual sense) [Fig.~{\ref{ImpedancePlane}}(a)]. At high current densities, the impedance changes from $\approx$$R_0$ to $Z^{1\omega}(\omega, j_0)$, of which imaginary component becomes a detectable magnitude. This nonlinearly induced change in the impedance is given by $\Delta Z^{1\omega}(\omega, j_0) \equiv Z^{1\omega}(\omega, j_0) - R_0$, which is represented by the red arrow in Fig.~{\ref{ImpedancePlane}}(a). Note that, for this vector, the real part is much greater than the imaginary part, and $\Delta Z^{1\omega}(\omega, j_0)$ has dissipative characteristics [Fig.~{\ref{ImpedancePlane}}(b)]. In contrast, in Refs.~\cite{YokouchiNature, KitaoriPNAS, YokouchiArxiv}, when considering the origin of the nonlinear impedance, the authors referred to $Z^{1\omega}(\omega, j_0) - Z^{1\omega}(\omega = 0, j_0)$, which is represented by the green arrow in Fig.~{\ref{ImpedancePlane}}(a), i.e., they focused on the frequency dependence of the nonlinear impedance, $Z^{1\omega}(\omega, j_0)$. For this green vector, the imaginary part is much greater than the real part [Fig.~{\ref{ImpedancePlane}}(b)], and by comparing this observation with that expected for a hypothetical negative inductor [Fig.~{\ref{ImpedancePlane}}(c)], the authors concluded that the nonlinearly induced impedance behaves like an inductor with negative inductance \cite{YokouchiNature, KitaoriPNAS, YokouchiArxiv}. However, it is evident that the nonlinearly induced impedance is represented by the red arrow, and therefore the consideration of the nonlinearly induced impedance by referring to the green arrow does not provide an accurate interpretation. Also note that a negative inductor is unstable, as discussed in Section~II~A and the Supplemental Material of Ref.~\cite{PrepriFuruta}.

\subsection{IV. Cross-sectional area dependence of the nonlinear impedance}
If we understand the argument in the comment \cite{Condmat} correctly, the authors seem to agree with our conclusion \textit{I} but hold a different opinion regarding our conclusion \textit{II}. Thus, the main issue to address is whether the resistor-like nonlinear impedance originates from Joule heating or other nonthermal intrinsic mechanisms. In the comment \cite{Condmat}, the authors claimed that the nonlinear impedance is ascribed to the nonthermal and intrinsic mechanism associated with the EEF \cite{Volovik, BarnesPRL2007, NagaosaJJAP} due to the STT-induced dynamics of the magnetic textures \cite{STT1, STT2} or spin fluctuations \cite{Fluctuations}. In this section, we show that, at least in Gd$_3$Ru$_4$Al$_{12}$, the sample geometry dependence of $\Im Z^{1\omega}$ observed in the original paper \cite{YokouchiNature} contradicts the authors' claim.

\begin{figure*}
\includegraphics{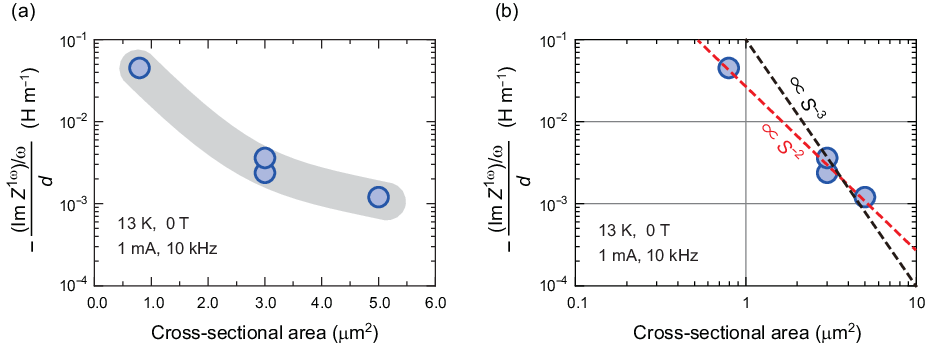}
\caption{\label{Nature_S_scaling} Cross-sectional area $S$ dependence of the nonlinear impedance in Gd$_3$Ru$_4$Al$_{12}$. (a) Original figure on a linear-log scale, reproduced from Ref.~\cite{YokouchiNature}. (b) The same data as in panel (a), replotted on a log--log scale.
}
\end{figure*}

Following the discussion in the first experimental report \cite{YokouchiNature}, we assume that, for the moment, the imaginary part of nonlinear complex resistivity, $\Im \rho^{1\omega}$, reflects the bulk properties and expand $\Im \rho^{1\omega}$ into a Taylor series in the amplitude $j_0$ of AC current density $j(t) = j_0 \sin \omega t$:
\begin{align}
\label{Taylor_j}
\Im \rho^{1\omega} &= {\rm c}_0 + {\rm c}_2j_0^2 +{\rm c}_4j_0^4 + \cdots,  
\end{align}
where c$_i$ are constants. To derive Eq.~(\ref{Taylor_j}), the authors \cite{Condmat} made the physically reasonable assumption that transport properties are symmetric with respect to current reversal; therefore, Eq.~(\ref{Taylor_j}) does not include terms of odd order in $j_0$. For a given current amplitude $I_0$, Eq.~(\ref{Taylor_j}) is rewritten as:
\begin{align}
\label{Taylor_I}
\Im Z^{1\omega} \frac{S}{d}= {\rm c}_0 + \frac{{\rm c}_2}{S^2}I_0^2 +\frac{{\rm c}_4}{S^4}I_0^4 + \cdots. \notag \\
\therefore \frac{\Im Z^{1\omega}}{d} = \frac{{\rm c}_0}{S} + \frac{{\rm c}_2}{S^3}I_0^2 +\frac{{\rm c}_4}{S^5}I_0^4 + \cdots,   
\end{align}
where $d$ and $S$ represent the distance between the voltage-probing electrodes and the sample cross-sectional area, respectively. In the previous experiments \cite{YokouchiNature, KitaoriPNAS, YokouchiArxiv}, $\Im Z^{1\omega}$ at low current densities was too small to be detected, as mentioned above, indicating that ${\rm c}_0$ is negligible in practice. Thus, if the authors' assumption, Eq.~(\ref{Taylor_j}), is correct, then the nonlinear $\Im Z^{1\omega} /d$ should scale with $1/S^3$ at a given $I_0$ [Eq.~(\ref{Taylor_I})]. In the first experimental report \cite{YokouchiNature}, the $S$ dependence of $\Im Z^{1\omega}$ was displayed on a linear-log scale, as reproduced in Fig.~{\ref{Nature_S_scaling}}(a). Figure {\ref{Nature_S_scaling}}(b) shows the $S$ dependence of $\Im Z^{1\omega} /d$ plotted on a log--log scale, revealing that $\Im Z^{1\omega} /d$ scales with $S^{-2}$ rather than $S^{-3}$. The observed nonlinear $\Im Z^{1\omega} /d$ contradicts the authors' assumption represented by Eqs.~(\ref{Taylor_j}) and (\ref{Taylor_I}). The behavior of $\Im Z^{1\omega} /d \propto S^{-2}$ leads us to conclude that the nonlinear $\Im Z^{1\omega}$ observed in Gd$_3$Ru$_4$Al$_{12}$ is at least not due to bulk properties. If the nonlinear impedance reflected the bulk intrinsic properties, such as the EEF, $\Im Z^{1\omega} /d$ should exhibit $S^{-3}$ scaling \cite{Note}.

Having established that the validity of interpreting the nonlinear impedance observed for Gd$_3$Ru$_4$Al$_{12}$ based on the EEF mechanism is negated by the observation of $\Im Z^{1\omega} \propto S^{-2}$ scaling, we can now ask if the Joule heating model can explain the $S^{-2}$ behavior. Because the contact resistance, $R_{\rm contact}$, and the thermal-response function, $\chi^{*}(\omega, T, H)$, which appear in the Joule heating model (see the equations in our preprint \cite{PrepriFuruta}), do not simply scale with $S$ and $d$, the $S$ dependence of $\Im Z^{1\omega}$ is not trivial. As shown below, a qualitative argument can be made by referring to the experimental results in a different material, namely, FeSn$_2$ \cite{YokouchiArxiv}. First, recall that for FeSn$_2$, the temperature dependence of $\Im Z^{1\omega}$ scales with that of $R_0\times ({\rm d}R_0/{\rm d}T)$ \cite{PrepriFuruta}, leading us to conclude that $\Im Z^{1\omega}$ in FeSn$_2$ \cite{YokouchiArxiv} is dominated by time-varying Joule heating, which occurs mainly in the bulk. In the experiments with FeSn$_2$, no obvious refutation of this explanation was found in the comment \cite{Condmat}. In Ref.~\cite{YokouchiArxiv}, Yokouchi \textit{et al}.~reported that the nonlinear $\Im Z^{1\omega} /d$ scales with $S^{-3}$ in FeSn$_2$. This observation is consistent with the nonlinear impedance due to Joule heating in the bulk. Given that our minimal Joule heating model predicts $\Im Z^{1\omega} \propto R_0\frac{{\rm d}R_0}{{\rm d}T}\chi''$ \cite{PrepriFuruta}, the observation of $\Im Z^{1\omega} \propto S^{-3}$ implies that
\begin{align}
\label{S^3}
\chi'' &\propto S^{-1}.   
\end{align}
We then discuss $\Im Z^{1\omega}$ in Gd$_3$Ru$_4$Al$_{12}$. In this material, the temperature dependence of $\Im Z^{1\omega}$ scales with that of ${\rm d}R_0/{\rm d}T$ \cite{PrepriFuruta}, implying that Joule heating occurs mainly at the contacts. In this case, the scaling of $S$ is given as:
\begin{align}
\label{S^2}
\Im Z^{1\omega} &\propto R_{\rm contact} \frac{{\rm d}R_0}{{\rm d}T}\chi'' \notag \\ 
\therefore \Im Z^{1\omega} &\propto R_{\rm contact} S^{-2}.   
\end{align}
Here, we use Eq.~(\ref{S^3}). Given that the electrodes are deposited mainly on the top surface of the microfabricated sample, the dependence of $R_{\rm contact}$ on $S$ might be weak.  Thus, under these assumptions, the behavior of $\Im Z^{1\omega} \propto S^{-2}$ in Gd$_3$Ru$_4$Al$_{12}$ [Fig.~{\ref{Nature_S_scaling}}(b)] may be explained by the Joule heating model. Furthermore, from the perspective of the Joule heating model, the different $S$-scaling behavior between Gd$_3$Ru$_4$Al$_{12}$ ($\Im Z^{1\omega} \propto S^{-2}$) and FeSn$_2$ ($\Im Z^{1\omega} \propto S^{-3}$) appears reasonable. This observation likely reflects the different ${\rm d}R_0/{\rm d}T$-scaling behavior between Gd$_3$Ru$_4$Al$_{12}$ ($\Im Z^{1\omega} \propto {\rm d}R_0/{\rm d}T$) and FeSn$_2$ [$\Im Z^{1\omega} \propto R_0 \times ({\rm d}R_0/{\rm d}T)$]. For more details on the data, refer to the figures in our preprint \cite{PrepriFuruta}. 

In summary, we find that in Gd$_3$Ru$_4$Al$_{12}$, $\Im Z^{1\omega}$ scales with $S^{-2}$, which is incompatible with Eqs.~(\ref{Taylor_j}) and ~(\ref{Taylor_I}). Thus, the behavior of $\Im Z^{1\omega} \propto S^{-2}$ becomes a critical issue when considering the origin of the nonlinear impedance in Gd$_3$Ru$_4$Al$_{12}$.  

\subsection{V. Scope of applicability of the minimal Joule heating model}
In the comment \cite{Condmat}, the authors argued that some of their data cannot be explained by the Joule heating model introduced in our preprint \cite{PrepriFuruta}, attributing the nonlinear impedance to the EEF instead. We are concerned that the authors may not fully recognize that the model in Ref.~\cite{PrepriFuruta} is minimal and that they perform incorrect data analysis for cases beyond the scope of minimal consideration. Here, we explain the scope of applicability of the model in more detail.

In our preprint \cite{PrepriFuruta}, we assumed that time-varying Joule heating is so small that $\partial R(T, H)/\partial T$ and the thermal response function $\chi^{*}(\omega, T, H)$ can be approximated as constants. In the Joule heating model in Ref.~\cite{PrepriFuruta}, we treated the effects that are linear with the input power amplitude $P_0(I_0, T, H)$, which is proportional to $I_0^2$; hereafter we refer to this model as the $P_0$-linear minimal Joule heating model to distinguish it from the Joule heating model including higher order $P_0^n$ terms ($n \geq 2$). When the variations of $\partial R(T, H)/\partial T$ and $\chi^{*}(\omega, T, H)$ within the considered Joule heating are not negligible, the temperature derivatives of the two quantities, $\partial^2R(T, H)/\partial T^2$ and $\partial \chi^*(\omega, T, H)/\partial T$, should also be accounted. To this end, we expand the higher order contributions to the Joule-heating-induced AC electrical response by considering that the unit of the impedance is ohm. For instance, beyond the $P_0$-linear minimal Joule heating model \cite{PrepriFuruta}, $\Re \Delta Z^{1\omega} \equiv \Re Z^{1\omega}(\omega, I_0, T, H) - R_0(T, H)$ is given by
\begin{align}
\label{HigherOrders}
\Re \Delta Z^{1\omega} &(\omega, I_0, T, H) \nonumber \\
=\: &{\rm A} \, \frac{\partial R_0}{\partial T} P_0(I_0, T, H) \, f\Big(\chi_0(T, H), \chi'(\omega, T, H), \chi''(\omega, T, H) \Big) \nonumber \\
&+ {\rm B} \, \frac{\partial^2 R_0}{\partial T^2} \Big( P_0(I_0, T, H) \Big)^2 \, g\Big(\chi_0(T, H), \chi'(\omega, T, H), \chi''(\omega, T, H)\Big) \nonumber \\
&+ {\rm C} \, \frac{\partial R_0}{\partial T} \Big( P_0(I_0, T, H) \Big)^2 \, h\bigg( \chi_0(T, H), \chi'(\omega, T, H), \chi''(\omega, T, H), \frac{\partial\chi_0}{\partial T}, \frac{\partial \chi'}{\partial T}, \frac{\partial\chi''}{\partial T} \bigg) \nonumber \\
&+ O\big( P_0^4 \big),
\end{align}
where A, B, and C represent real coefficients, the function $f$ has the unit of 
[K~W$^{-1}$] and is expressed in the linear form of the arguments. The function $g$ has the unit of [K$^2$~W$^{-2}$] and is written in the quadratic form of the arguments, while the function $h$ has the unit of [K~W$^{-2}$] and is written in the quadratic form of the product of the nondifferential and differential arguments such as $\chi_0 \frac{\partial\chi'}{\partial T}$ and $\chi' \frac{\partial \chi'}{\partial T}$. Similar expansions are applied to $\Im Z^{1\omega}$, $\Re Z^{3\omega}$, and $\Im Z^{3\omega}$. As discussed in our preprint \cite{PrepriFuruta}, only when the contributions from these higher-order terms are negligible can we rely on the $P_0$-linear minimal Joule heating model [i.e., the first term on the right-hand side of Eq.~(\ref{HigherOrders})]. The following clear conclusions can be drawn in this case \cite{PrepriFuruta}. First, the signs of $\Re \Delta Z^{1\omega}$, $\Im Z^{1\omega}$, $\Re Z^{3\omega}$, and $\Im Z^{3\omega}$ are determined by the sign of $\partial R_0/\partial T$ (Table I). Second, as $\Re \Delta Z^{1\omega}(\omega \rightarrow \infty)/ \Re \Delta Z^{1\omega}(\omega = 0) = 2/3$ because $f = 2\chi_0(\omega, T, H) + \chi'(\omega, T, H)$ for $\Re \Delta Z^{1\omega}$. Third, $\Im Z^{3\omega}(\omega, I_0, T, H)/\Re Z^{3\omega}(\omega, I_0, T, H) = -\chi''(\omega, T,H)/\chi'(\omega, T, H)$, which is negative and independent of $I_0$. These simplified outcomes are expected only when Joule heating is so small that the variations in $\partial R(T, H)/\partial T$ and $\chi^{*}(\omega, T, H)$ are negligible. As the system moves into the highly nonlinear regime, the above outcomes derived from the $P_0$-linear minimal Joule heating model must be modified. It is important to note that deviations from this model do not necessarily suggest that nonthermal mechanisms, such as the EEF, play a dominant role in the nonlinear impedance.

\begin{figure*}[b]
\includegraphics{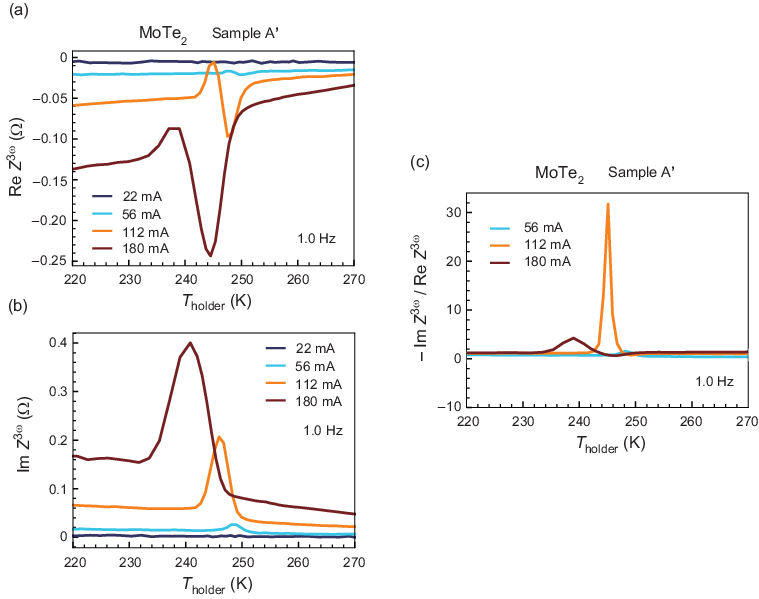}
\caption{\label{Ratio} Temperature dependence of (a) $\Re Z^{3\omega}$, (b) $\Im Z^{3\omega}$, and (c) $-\Im Z^{3\omega} / \Re Z^{3\omega}$ in MoTe$_2$ (Sample A' in our study \cite{PrepriFuruta}).
}
\end{figure*}

When comparing the $P_0$-linear minimal Joule heating model to the experimental results, attention should be paid to its scope of applicability. Generally, this model becomes less sufficient at and near phase transition points. For instance, issues associated with latent heat fall outside the scope of Eq.~(\ref{HigherOrders}). Furthermore, even without latent heat, $\partial R(T, H)/\partial T$ and $\chi^*(\omega, T, H)$ may change more notably in the critical region for a given amount of Joule heat. The authors \cite{Condmat} appeared to overlook this point. In fact, the authors \cite{Condmat} seem to suggest that the peak anomaly in the ratio of $\Im Z^{3\omega}$ to $\Re Z^{3\omega}$ at the magnetic phase transition should be attributed to the EEF mechanism because it is not entirely explained by the $P_0$-linear minimal Joule heating model. We discuss this issue in Section~VI in more detail.

A quantitative understanding of the nonlinear impedance in a highly nonlinear regime is outside the scope of this paper, and its physical significance in exploring the origin of nonlinear impedance is also not clear. Nevertheless, to address this, we examine the results from MoTe$_2$ (Sample A' in our preprint \cite{PrepriFuruta}) in more detail and present data showing deviations from the $P_0$-linear minimal Joule heating model at and near phase transition points. Figure {\ref{Ratio}}(a)--(c) shows the temperature dependence of $\Re Z^{3\omega}$, $\Im Z^{3\omega}$, and the ratio of $\Im Z^{3\omega}$ to $\Re Z^{3\omega}$, respectively, near the structural-transition temperature. In Fig.~{\ref{Ratio}}(c), $-\Im Z^{3\omega} / \Re Z^{3\omega}$ exhibits peak anomalies at the structural phase transition. Our results demonstrate that $-\Im Z^{3\omega} / \Re Z^{3\omega}$ may be susceptible to singularities accompanying the phase transition. A closer examination of the data further reveals a nontrivial behavior of the nonlinear impedance at and near the phase transition. In Fig.~{\ref{Ratio}}(a), $\Re Z^{3\omega}$ exhibits nonmonotonous (resonance-like in shape) anomalies, as seen most prominently in the data at 112 mA. Furthermore, whereas the $\Im Z^{3\omega}$ increases as the current increases and is most pronounced at the maximum current, 180 mA [Fig.~{\ref{Ratio}}(b)], the anomaly in $-\Im Z^{3\omega} / \Re Z^{3\omega}$ is most pronounced at the intermediate current, 112 mA [Fig.~{\ref{Ratio}}(c)]. These observations cannot be directly explained within the framework of the $P_0$-linear minimal Joule heating model. However, given that the nonlinear impedance in MoTe$_2$ is generally well described by the minimal Joule heating model \cite{PrepriFuruta}, it is reasonable to assume that the nonlinear impedance observed at the phase transition is also caused by the time-varying Joule heating, including pronounced contributions of higher order $P_0^n$ terms in Eq.~(\ref{HigherOrders}). 

In summary, even though deviations from the $P_0$-linear minimal Joule heating model are observed at and near the phase transition, the probable effect of Joule heating should still be considered. A peak anomaly in $-\Im Z^{3\omega} / \Re Z^{3\omega}$ at the phase transition is not necessarily related to enhanced EEF.


\subsection{VI. Comment on each figure presented in the comment \cite{Condmat}}

\subsection{A. General remarks} 
Figures 1--4 in the comment \cite{Condmat} focus on discussing the transport properties of Gd$_3$Ru$_4$Al$_{12}$. The authors argued that the newly presented figures demonstrate that the observed nonlinear resistive impedance originates from the AC-current-induced dynamics of the noncollinear magnetic textures via the STT effect. However, we interpret the implications of these newly presented data differently. To correctly understand Figs.~1--4 in the comment \cite{Condmat}, the following remarks should be noted. 

First, the $\Im Z^{1\omega} \propto S^{-2}$ behavior (Fig.~{\ref{Nature_S_scaling}}) is inconsistent with the EEF-based scenario [Eq.~({\ref{Taylor_I}})]. Thus, there is no longer any basis for interpreting the data in terms of the EEF, at least for Gd$_3$Ru$_4$Al$_{12}$. 

Second, when considering the origin of the nonlinear impedance, the real part becomes more important aspect to note, since the nonlinear impedance observed in Gd$_3$Ru$_4$Al$_{12}$ \cite{YokouchiNature} has been identified as resistive in nature \cite{PrepriFuruta}. However, when considering the dynamics of the nonlinear impedance in more detail, it is important to also consider the imaginary component.

\begin{figure*}
\includegraphics{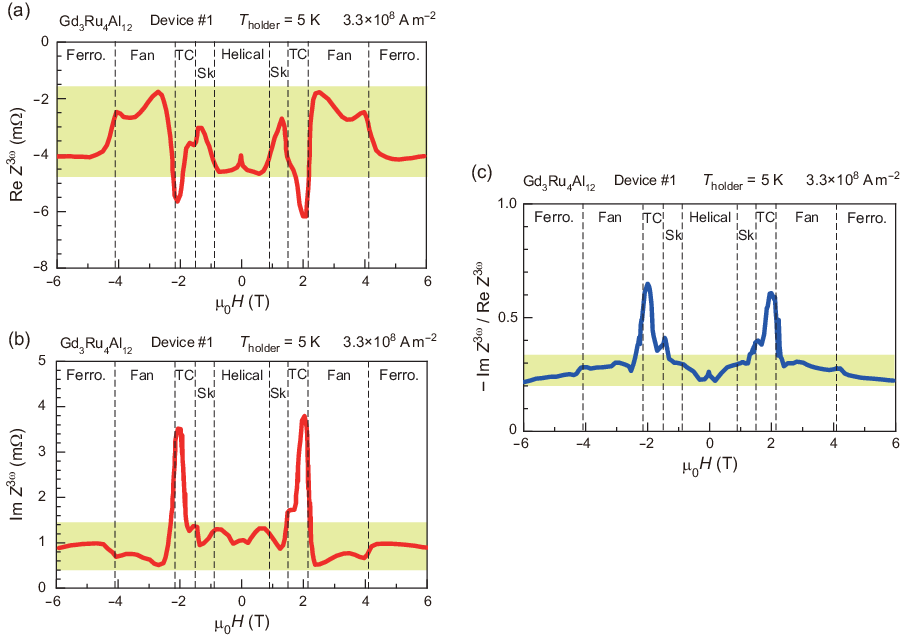}
\caption{\label{H_dep} Magnetic field dependence of (a) $\Re Z^{3\omega}$, (b) $\Im Z^{3\omega}$ and (c) $-\Im Z^{3\omega} / \Re Z^{3\omega}$ for the microfabricated Gd$_3$Ru$_4$Al$_{12}$ device at $T_{\rm holder} = 5$ K. The current density is 3.3$\times$10$^8$ A~m$^{-2}$. Note that as shown in Fig.~{\ref{SC_contami}}, the device is under the influence of extrinsic superconductivity (SC). Panel (a) is reproduced from Fig.~3(i)--(n) in the comment \cite{Condmat}, and $-\Im Z^{3\omega}/\Re Z^{3\omega}$ is calculated based on (a) and (b).
}
\end{figure*}

Third, in the comment \cite{Condmat}, the nonlinear impedance is observed also in the ferromagnetic phase, with a magnitude comparable to that observed in other noncollinear magnetic phases, such as the helical, skyrmion (Sk), and fan phases. For clarity, we reproduce the representative magnetic-field dependence of $\Re Z^{3\omega}$, $\Im Z^{3\omega}$, and $-\Im Z^{3\omega}/\Re Z^{3\omega}$, as shown in Fig.~{\ref{H_dep}}(a)--(c), respectively. Although the magnetic field dependence exhibits rich structures, the nonlinear impedance (both the real and imaginary parts) remains on the same order throughout the field sweep, from the helical phase at zero field to the ferromagnetic phase above 4 T. Nonlinear impedance that is insensitive to magnetic textures implies that the noncollinearity in the ordered (i.e., time-averaged) magnetic texture is not a crucial factor in the emergence of the nonlinear impedance. The authors \cite{Condmat} claimed that the signal observed in the ferromagnetic phase is mainly due to the spin fluctuations around the time-averaged structure \cite{Fluctuations}, whereas the signals observed in the presence of noncollinear magnetic textures are mainly due to the current-induced dynamics of magnetic textures, aside from the fact that both scenarios are inconsistent with the $\Im Z^{1\omega} \propto S^{-2}$ behavior. To explain the signals that remain on the same order throughout the field sweep, we propose that the results are more likely attributable to a single mechanism, that is, time-varying Joule heating \cite{PrepriFuruta}.

\begin{figure*}
\includegraphics{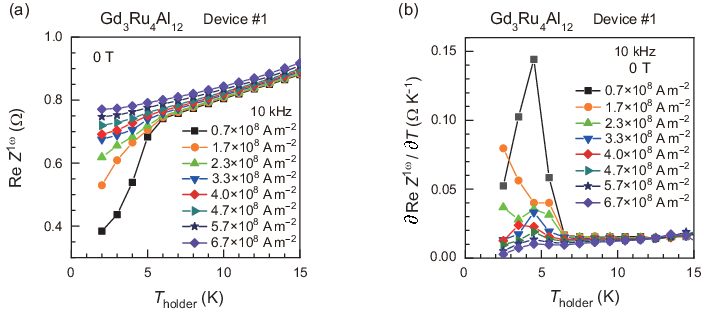}
\caption{\label{SC_contami}  Temperature dependence of (a) $\Re Z^{1\omega}$ and (b) $\frac{\partial}{{\partial}T}\Re Z^{1\omega}$ for the Gd$_3$Ru$_4$Al$_{12}$ microfabricated device contaminated by extrinsic superconductivity measured under various current densities. 
}
\end{figure*}

Fourth, and most importantly, the data on Gd$_3$Ru$_4$Al$_{12}$ presented in the comment \cite{Condmat} are affected by extrinsic superconductivity (SC) introduced during the microfabrication process using a focused ion beam (FIB). The authors used gallium and tungsten in device fabrication. However, it is well known that extrinsic SC consisting of tungsten, gallium, and carbon is often formed \cite{ContamiSC}. The critical temperature of this extrinsic SC is 5.2 K \cite{ContamiSC}. The data on Gd$_3$Ru$_4$Al$_{12}$ presented in the comment \cite{Condmat} were collected at 5 K. Therefore, the transport characteristics presented in the comment \cite{Condmat} are likely under the influence of extrinsic SC. Note that Gd$_3$Ru$_4$Al$_{12}$ is originally not a superconductor \cite{Gd3Ru4Al12}, but the $\Re Z^{1\omega}$--$T$ profile of the Gd$_3$Ru$_4$Al$_{12}$ device under consideration exhibits a pronounced decrease below 6 K under the lowest current density (0.7$\times$10$^8$ A~m$^{-2}$), whereas the decrease is much less pronounced under large current densities, i.e., above 2.3$\times$10$^8$ A~m$^{-2}$ [Fig.~{\ref{SC_contami}}(a)]. It follows that $\frac{\partial}{{\partial}T}\Re Z^{1\omega}$ significantly varies with temperature and, more importantly, with current density [Fig.~{\ref{SC_contami}}(b)]. Thus, the low-field and low-current-density data collected at 5 K clearly fall outside the scope of the $P_0$-linear minimal Joule heating model, and deviations are reasonable. Obviously, at 5 K, $\frac{\partial}{{\partial}T}(\frac{\partial}{{\partial}T}\Re Z^{1\omega})$ is large and negative when the impact of the extrinsic SC is present [Fig.~{\ref{SC_contami}}(b)]; otherwise, it is small and positive. According to the literature \cite{ContamiSC}, an out-of-plane magnetic field higher than 1 T is needed to completely suppress the effect of the extrinsic SC at 5 K. This extrinsic SC is not discussed in the comment \cite{Condmat}, and, therefore, the authors could not elaborate on the systematics in the data. However, as discussed below, considering this extrinsic SC provides a qualitative understanding of the systematics in the data. 

Finally, as noted in the comment \cite{Condmat}, the $P_0$-linear Joule heating model does not directly explain the enhancement of nonlinear $\Im Z^{3\omega}$ at and near the phase transition point ($\approx$2 T) separating the transverse conical (TC) and fan phases. As discussed in Section~V, deviations from the $P_0$-linear minimal Joule heating model are reasonable, and they do not support the EEF mechanism, nor do they exclude the Joule heating scenario. 

In the following sections, we comment in more detail on each of the figures in the comment \cite{Condmat}. Since the EEF-based scenario is refuted by the observation of $\Im Z^{1\omega} \propto S^{-2}$ (Fig.~{\ref{Nature_S_scaling}}), the main focus here is whether the new data from the authors \cite{Condmat} truly refute the Joule-heating-based scenario, especially for Gd$_3$Ru$_4$Al$_{12}$. To gain a comprehensive understanding of the nonlinear impedance of this material, we first address Fig.~3 in Ref.~\cite{Condmat}, which displays the magnetic field dependence at various current densities. Then, Figs.~4, 1, and 2 are reviewed. Note that Subsections B--E can be read independently. Please understand that duplicate comments may appear. In Subsection F, we briefly discuss the results for Y$_{1-x}$Tb$_{x}$Mn$_6$Sn$_6$.

\subsection{B. Remarks on Fig.~3 in the comment \cite{Condmat}}

\begin{figure*}
\includegraphics{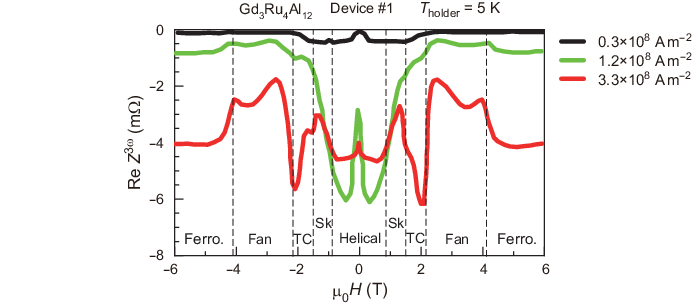}
\caption{\label{Current_dep} $\Re Z^{3\omega}$--$H$ profile of the Gd$_3$Ru$_4$Al$_{12}$ microfabricated device contaminated by extrinsic superconductivity at selected current densities. The data are reproduced from Fig.~3 in the comment \cite{Condmat}.
}
\end{figure*}
Figure 3 in Ref.~\cite{Condmat} displays the magnetic field dependence at various current densities. For clarity, we display the $\Re Z^{3\omega}$--$H$ profile at $T_{\rm holder} = 5$ K and selected current densities in Fig.~{\ref{Current_dep}}. As $j_0$ increases, the nonlinear impedance below 1 T exhibits nonmonotonous changes (for instance, the signal of the largest magnitude is observed at the intermediate current density), whereas the data above 2 T exhibits systematic increases in magnitude. Considering the impact of the extrinsic SC at low fields and low current densities, the $\Re Z^{3\omega}$ variation with $j_0$ can be understood, at least qualitatively, within the framework of the Joule-heating-based scenario. The conclusion is twofold. First, for data above 2 T, the impact of the extrinsic SC is negligible, and thus, $\Re Z^{3\omega}$, including in the ferromagnetic phase, monotonically increases as the Joule-heating-induced temperature oscillation becomes pronounced. Second, for data below 1 T, the impact of the extrinsic SC should be considered. Notably, $\frac{\partial}{{\partial}T}\Re Z^{1\omega}$ at $T_{\rm holder} = 5$ K steeply decreases as $j_0$ increases [Fig.~{\ref{SC_contami}}(b)]; thus, with increasing $j_0$, $\Re Z^{3\omega}$ may exhibit nonmonotonous changes due to the competition between decreasing $\frac{\partial}{{\partial}T}\Re Z^{1\omega}$ and increasing temperature oscillation. From another perspective, the significant difference in magnitude between the high- and low-field data should only be characteristic for the data at low current densities, such as 1.2$\times$10$^8$ A~m$^{-2}$. For high current densities, in contrast, the extrinsic SC at $T_{\rm holder} = 5$ K is suppressed [Fig.~{\ref{SC_contami}}(a)]. Thus, $\frac{\partial}{{\partial}T}\Re Z^{1\omega}$ is expected to remain on the same order throughout the field sweep; as a result, $\Re Z^{3\omega}$ also remains on the same order, regardless of the magnetic phases (Fig.~{\ref{Current_dep}}). In addition, in Fig.~{\ref{Current_dep}}, the anomaly associated with the helical--skyrmion transition at $\approx$0.9 T is not clearly visible for 1.2$\times$10$^8$ A~m$^{-2}$ because the steep suppression of the SC under increasing magnetic field masks the anomaly. In contrast, the impact of the extrinsic SC is suppressed under 3.3$\times$10$^8$ A~m$^{-2}$ at $T_{\rm holder} = 5$ K [Fig.~{\ref{SC_contami}}(a)], and thus, the transition anomaly becomes visible in the $\Re Z^{3\omega}$--$H$ profile measured under 3.3$\times$10$^8$ A~m$^{-2}$. 

The authors in the comment \cite{Condmat} argued that Joule heating has negligible effect for current densities below 1.7$\times$10$^{8}$ A~m$^{-2}$; for instance, the authors' analysis suggests that the average temperature increase $\Delta T$ is much smaller than 0.1 K for 1.7$\times$10$^{8}$ A~m$^{-2}$. If this estimation is correct, the nonlinear impedance observed at and below 1.7$\times$10$^{8}$ A~m$^{-2}$ is not due to time-varying Joule heating, as the authors claimed \cite{Condmat}. Therefore, we examine whether $\Delta T$ is truly negligible below 1.7$\times$10$^{8}$ A~m$^{-2}$. At a current density of 3.3$\times$10$^{8}$ A~m$^{-2}$, the authors argued that $\Delta T \approx 1.1$ K. Given that the power is proportional to $j^2$, we expect that $\Delta T \approx 0.1$ K for 1.2$\times$10$^{8}$ A~m$^{-2}$. This estimation sharply contrasts that made by the authors. Thus, we question the results of the authors' analysis and verify the data used for the estimation of $\Delta T$.

\begin{figure*}
\includegraphics{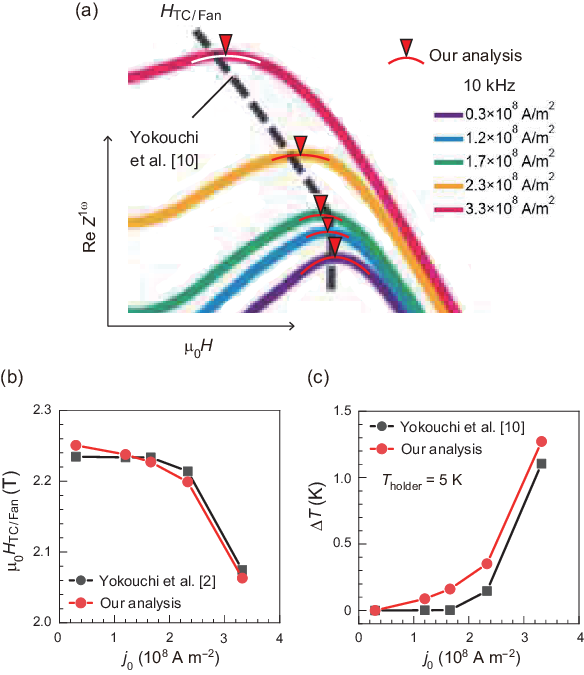}
\caption{\label{Copy} (a) Enlarged view of the $\Re Z^{1\omega}$--$H$ profile reproduced from Fig.~3 in the comment \cite{Condmat}. (b) The transition field $H_{\rm TC/Fan}$ between the transverse conical (TC) and fan magnetic phases, determined from $\frac{\partial}{{\partial}H}\Re Z^{1\omega} =0$ in panel (a). (c) Estimated average temperature increase $\Delta T$ due to Joule heating. The procedure followed is the same as that described in the comment \cite{Condmat}.
}
\end{figure*}

In the comment \cite{Condmat}, the authors examined the current-density dependence of the transition field between the TC and fan magnetic phases, $H_{\rm TC/Fan}$, at which $\frac{\partial}{{\partial}H}\Re Z^{1\omega} =0$. Then, they estimated the average temperature increase due to Joule heating from the current-induced-change of $H_{\rm TC/Fan}$. For reference, in Fig.~{\ref{Copy}}(a), we reproduce the $\Re Z^{1\omega}$--$H$ profile displayed in the comment \cite{Condmat}. By referring to these data, we use parabolic fittings to determine the $j_0$-dependent $H_{\rm TC/Fan}$. We then find that current-induced variations of $H_{\rm TC/Fan}$ are of a detectable magnitude, even at low current densities below 1.7$\times$10$^{8}$ A~m$^{-2}$ [Fig.~{\ref{Copy}}(a)]; in contrast, in the comment \cite{Condmat}, the authors concluded that there were no variations in $H_{\rm TC/Fan}$ for current densities below 1.7$\times$10$^{8}$ A~m$^{-2}$ [Fig.~{\ref{Copy}}(b)]. Thus, by referring to the temperature dependence of $H_{\rm TC/Fan}$, we estimate $\Delta T$, as shown in Fig.~{\ref{Copy}}(c). According to our analysis, $\Delta T$ is $\approx$ 0.09 K for 1.2$\times$10$^{8}$ A~m$^{-2}$, consistent with the expectation mentioned in the previous paragraph.

To verify the relevance of Joule heating, we perform an order-of-magnitude estimate of $\Re Z^{3\omega}$ within the framework of the $P_0$-linear minimal Joule heating model. To this end, we avoid the data at 1.2$\times$10$^{8}$ A~m$^{-2}$ and low fields because Fig.~{\ref{SC_contami}}(b) indicates that such data are outside the scope of this model. Instead, we focus on data above 3 T, which should be unaffected by the extrinsic SC, making the application of the $P_0$-linear minimal Joule heating model more reasonable. This model predicts $\Re Z^{3\omega} \approx -({\partial}R_0/{\partial}T)(\Delta T)/2$, where ${\partial}R_0/{\partial}T$ represents the value under negligible Joule heating \cite{PrepriFuruta}. $\Delta T$ is $\approx$ 0.09 K, and from Fig.~{\ref{SC_contami}}(b), the value of ${\partial}R_0/{\partial}T$ is expected to be on the order of 0.01 $\Omega$~K$^{-1}$ when the extrinsic SC is suppressed by the field. Thus, we obtain $\Re Z^{3\omega} \approx -0.4$ m$\Omega$, which is on the same order of magnitude as the actual observed value, i.e., between $-0.5$ and $-1$ m$\Omega$ (Fig.~{\ref{Current_dep}}). This order-of-magnitude agreement provides strong evidence that $\Re Z^{3\omega}$ includes a considerable contribution of the Joule-heating-induced electrical response, even at 1.2$\times$10$^{8}$ A~m$^{-2}$.

Finally, apart from the fact that the EEF-based scenario is found to be inconsistent with the $\Im Z^{1\omega} \propto S^{-2}$ behavior (Fig.~{\ref{Nature_S_scaling}}), we address the authors' discussion of Fig.~3 in the comment \cite{Condmat}. The authors \cite{Condmat} attributed the observed nonlinear impedance to the current-induced dynamics of the magnetic texture when a noncollinear magnetic texture is present or to spin fluctuations when a magnetic texture is absent. However, the authors provide no explanation as to (i) the validity of the magnitude of the observed nonlinear impedance, (ii) the difference of the systematics of the nonlinear impedance under the $j_0$ variations between the low and high fields, and (iii) the expected temperature and magnetic-field dependence of the nonlinear impedance.

\subsection{C. Remarks on Fig.~1 in the comment \cite{Condmat}}
\begin{figure*}[b]
\includegraphics{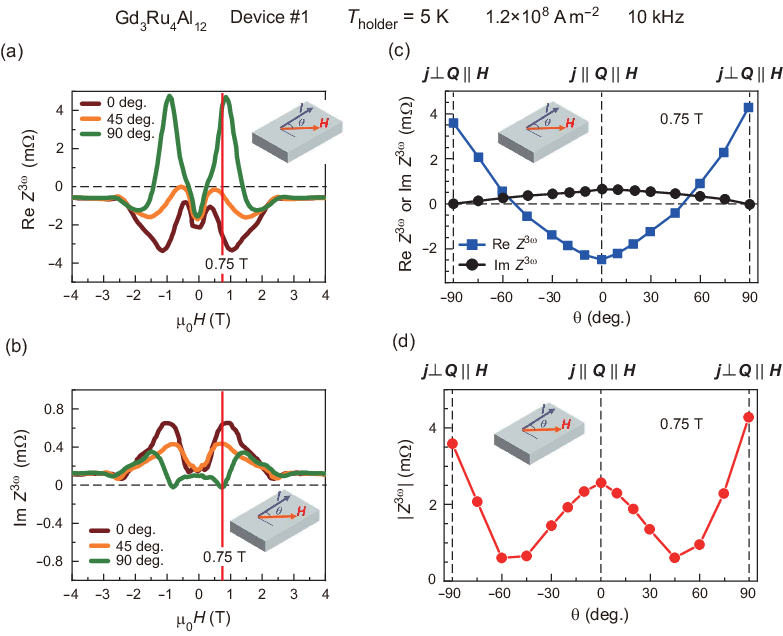}
\caption{\label{Angle} (a, b) Magnetic field dependence of $\Re Z^{3\omega}$ (a) and $\Im Z^{3\omega}$ (b) under selected in-plane-field angles. (c, d) In-plane-field angle dependence of (c) $\Re Z^{3\omega}$ and $\Im Z^{3\omega}$ and (d) $| Z^{3\omega} |$ at $T_{\rm holder} = 5$ K, measured with 1.2$\times$10$^8$ A~m$^{-2}$. The magnetic field is fixed at 0.75 T [indicated by the red solid line in (a, b)], which is close to the phase transition point between the multi domain helical and conical phases. Panels (a)--(c) are reconstructed from Fig.~1 in the comment \cite{Condmat}, and panel (d) is calculated based on panel (c).
}
\end{figure*}

The authors' conclusion that the nonlinear impedance should be due to the EEF mechanism is based partly on the results of the in-plane-field angle dependence of the nonlinear impedance (Fig.~1 in the comment \cite{Condmat}). Our careful examination of these results reveals that the angle-dependent data rather refutes the interpretation based on the EEF scenario.

Figure 1 in the comment \cite{Condmat} displays the nonlinear impedance ($\Re Z^{3\omega}$ and $\Im Z^{3\omega}$) at $T_{\rm holder} = 5$ K, focusing on the magnetic field dependence of under various in-plane-field angles, $\theta$. For clarity, we reproduce the data in Fig.~{\ref{Angle}}(a) and (b). For the field angle dependence at 0.75 T, we reproduce the data in Fig.~{\ref{Angle}}(c). Apart from the fact that the EEF-based scenario is found to be inconsistent with the $\Im Z^{1\omega} \propto S^{-2}$ behavior (Fig.~{\ref{Nature_S_scaling}}), the authors claimed that (i) the low-field and field-rotation data are inconsistent with the $P_0$-linear minimal Joule heating model and (ii) at 0.75 T, $\Im Z^{3\omega}$ systematically decreases from its maximum for the ${\bm j} \parallel {\bm Q}$ configuration (${\bm j}$ and ${\bm Q}$ represent the current-density vector and the magnetic modulation vector, respectively) to $\Im Z^{3\omega} \approx 0$ for the ${\bm j} \perp {\bm Q}$ configuration [Fig.~{\ref{Angle}}(b) and (c)], consistent with the scenario of the magnetic-texture dynamics induced by the STT effect \cite{STT1, STT2}. Note that the current-induced EEF mechanism, which is based on the STT effect, is most effective for the ${\bm j} \parallel {\bm Q}$ configuration. Conversely, for the ${\bm j} \perp {\bm Q}$ configuration, the STT effect does not apply, resulting in (ideally) no current-induced dynamics of magnetic textures and thus no EEF.

Regarding observation (i), the authors stated, for instance, that ``the angle dependence of ${\rm d}R/{\rm d}T$ is different from either $\Re Z^{3\omega}$ or $\Im Z^{3\omega}$'', thereby claiming that the data cannot be explained by the $P_0$-linear minimal Joule heating model. As discussed in Sections~V and ~VI A, the deviations from this model are rather reasonable. Again, the low-field data measured at low-current density, 1.2$\times$10$^8$ A~m$^{-2}$, likely reflect the influence of contamination by the extrinsic SC. Furthermore, the singularity near 0.75--1 T is associated with the transition between the conical and helical phases, further invalidating the application of the $P_0$-linear minimal Joule heating model. Therefore, the deviations from this model in the low-field and field-rotation data are not evidence that time-varying Joule heating is irrelevant to the nonlinear impedance.

The authors' interpretation of the observation (ii) seems to involve a serious misunderstanding. The magnetic field of 0.75 T is close to the transition field between the multidomain helical and conical phases, which implies that the general argument cannot be immediately derived from the field-rotation data at such a specific field. For instance, $\Im Z^{3\omega}$ at 0 T is rather insensitive to the in-plane-field direction, which contradicts the authors' conclusion derived from observation (ii). Nevertheless, we continue the discussion on the data at 0.75 T, as the authors \cite{Condmat} seemed to place great importance on these data. We show the field-angle $\theta$ dependence of $|Z^{3\omega}|$ in Fig.~{\ref{Angle}}(d). It is obvious that the magnitude of the nonlinear impedance becomes maximum for the ${\bm j} \perp {\bm Q}$ configuration, in which the STT effect does not apply. This observation contradicts the authors' claim that the observed nonlinear impedance originates from the current-induced dynamics of the magnetic texture via the STT effect. The authors' misinterpretation stems from the fact that they attempted to interpret the nonlinear impedance by focusing only on the imaginary part, as previously discussed in Section~III and Fig.~{\ref{ImpedancePlane}}. Moreover, as discussed in Section~II (Table I), the current-induced dynamics cannot induce positive $\Re Z^{3\omega}$, and note again that the STT-based scenario is refuted by the observation of the $\Im Z^{1\omega} \propto S^{-2}$ behavior (Fig.~{\ref{Nature_S_scaling}}).

The data beyond 2 T are insensitive to the in-plane-field direction, as also noted by the authors \cite{Condmat}. To explain the signals that remain on the same order throughout the magnetic-field sweep, it is more appropriate to consider a single mechanism, which is, in our view, the Joule-heating-induced AC electrical response. Note again that as discussed in Section~VI B, the $P_0$-linear minimal Joule heating model predicts $\Re Z^{3\omega} \approx -0.4$ m$\Omega$ for 1.2$\times$10$^8$ A~m$^{-2}$ when the impact of the extrinsic SC is negligible, and this result agrees with the actual observed value [Fig.~{\ref{Angle}}(a)]. In contrast, the EEF-based scenario does not currently offer the order of magnitude of the expected nonlinear impedance. Furthermore, a good correlation between the $\Im \rho^{1\omega}$--$H$ and $\frac{\partial}{{\partial}T} \Re \rho^{1\omega}$--$H$ profiles is identified at $T_{\rm holder} = 5$ K and 3.3$\times$10$^8$ A~m$^{-2}$ throughout the field sweep \cite{PrepriFuruta}. At this current density, the impact of the extrinsic SC is suppressed, even at 0 T [Fig.~{\ref{SC_contami}}(a) and (b)], and the complexity due to the extrinsic SC diminishes from the field-dependent measurements. The similarity between the $\Im \rho^{1\omega}$--$H$ and $\frac{\partial}{{\partial}T} \Re \rho^{1\omega}$--$H$ profiles is only qualitative, but the EEF-based scenario has never provided a shape of the expected $\Im \rho^{1\omega}$--$H$ profile even at a qualitative level.

\subsection{D. Remarks on Fig.~4 in the comment \cite{Condmat}}
Figure 4 in the comment \cite{Condmat} presents the magnetic field dependence of $-\Im Z^{3\omega}/\Re Z^{3\omega}$ at $T_{\rm holder} = 5$ K, which is essentially the same as our Fig.~{\ref{H_dep}}(c). Regarding the singular behavior at the phase transition point ($\approx$2 T), the authors stated that, ``This correlation between $-\Im Z^{3\omega}/\Re Z^{3\omega}$ and the magnetic phase indicates that $\Re Z^{3\omega}$ and $\Im Z^{3\omega}$ are related spin textures'' \cite{Condmat}. However, the STT-based scenario is refuted by the observation of the $\Im Z^{1\omega} \propto S^{-2}$ behavior (Fig.~{\ref{Nature_S_scaling}}) and also by the maximum signal for the ${\bm j} \perp {\bm Q}$ configuration [Fig.~{\ref{Angle}}(d)]. The authors claimed that the anomaly at the phase transition cannot be explained by Joule heating. As discussed in Section~V, deviations from the $P_0$-linear minimal Joule-heating model are reasonably expected for data at and near a phase transition, and therefore, such deviations do not indicate that Joule heating is irrelevant to the nonlinear impedance at the phase transition point. Furthermore, as mentioned in Section~VI A, $-\Im Z^{3\omega}/\Re Z^{3\omega}$ remains on the same order, regardless of the magnetic phases, including the ferromagnetic phases above 4 T [Fig.~{\ref{H_dep}(c)]. A straightforward implication of this observation is that a magnetic-phase-independent mechanism dominates the overall nonlinear impedance characteristics, with the most probable and simplest scenario being time-varying Joule heating caused by a large AC current.

\subsection{E. Remarks on Fig.~2 in the comment \cite{Condmat}}
Figure 2 in the comment \cite{Condmat} presents the frequency dependence of the nonlinear impedance under various current densities at $T_{\rm holder} = 5$ K and 0 T. These data do not show the relationship $\Re \Delta Z^{1\omega}(\omega \rightarrow \infty)/ \Re \Delta Z^{1\omega}(\omega = 0) = 2/3$, which is a consequence of the $P_0$-linear minimal Joule heating model. The authors argued that the data should therefore originate not from Joule heating but from the current-induced dynamics of the magnetic textures. Again, the STT-based scenario is refuted by the observation of the $\Im Z^{1\omega} \propto S^{-2}$ behavior (Fig.~{\ref{Nature_S_scaling}}) and the maximum signal for the ${\bm j} \perp {\bm Q}$ configuration (Fig.~{\ref{Angle}}). Therefore, the question to consider here is whether the deviation from the $P_0$-linear minimal Joule heating model is reasonable under the measurement conditions. As mentioned above, the data presented by the authors \cite{Condmat} are measured at $T_{\rm holder} = 5$ K and 0 T, and thus, the data should be affected by the extrinsic SC [Fig.~{\ref{SC_contami}}(a) and (b)]. Thus, as discussed in Sections~V and VI A, deviations from the minimal Joule-heating model are reasonable and do not indicate that Joule heating is irrelevant.

\subsection{F. Summary for Gd$_3$Ru$_4$Al$_{12}$}

The authors \cite{Condmat} claimed that the observed nonlinear impedance originates from the current-induced dynamics of the magnetic texture when a noncollinear magnetic texture is present and from spin fluctuations \cite{Fluctuations} when a magnetic texture is absent. However, the observation that $\Im Z^{1\omega}$ scales with $S^{-2}$ does not support either scenario. Furthermore, the in-plane-field angle dependence of the nonlinear impedance at 0.75 T exhibits the largest magnitude for the ${\bm j} \perp {\bm Q}$ configuration, and this observation is inconsistent with the EEF due to the STT-induced dynamics of the magnetic texture. Therefore, discussing the data in the light of the EEF mechanism is unfounded, unless a new scenario is considered where the EEF scales with $S^{-2}$ and becomes maximum for the ${\bm j} \perp {\bm Q}$ configuration.

Thus, a discussion at this stage should focus on whether various data are beyond the expectation from the Joule-heating-induced AC electrical response. The authors referred to data collected at 5 K, which are affected by the extrinsic SC, and, in some cases, the data at and near a phase transition. They also found deviations from the $P_0$-linear minimal Joule heating model and concluded that Joule heating is irrelevant to the observed nonlinear impedance \cite{Condmat}. However, it is reasonably expected that under the influence of the extrinsic SC and the singularities associated with a phase transition, the $P_0$-linear minimal Joule heating model becomes less sufficient for describing the nonlinear impedance. Thus, the authors' findings do not at all refute the possibility that the nonlinear impedance is caused by time-varying Joule heating.

By referring to the scenario based on Joule heating, we provide a qualitative explanation for (i) why the nonlinear impedance remains on the same order throughout the magnetic field sweep, from the helical phase to the ferromagnetic phase, and (ii) the systematics of the $j_0$-dependent and field-dependent data under the influence of the extrinsic SC. Moreover, (iii) we provide an order of magnitude estimate of the nonlinear impedance and found that it is consistent with the observed value. In contrast, apart from the fact that the EEF-based scenario is ruled out by the $\Im Z^{1\omega} \propto S^{-2}$ behavior and the in-plane-field angle dependence, the original paper \cite{YokouchiNature} and comment \cite{Condmat} do not offer further discussion to compare with our (i)--(iii). At this stage, there is no evidence to positively support the EEF scenario for Gd$_3$Ru$_4$Al$_{12}$; in fact, the available data \cite{PrepriFuruta, YokouchiNature, Condmat} lead toward a negative conclusion.

\subsection{G. Remarks on Fig.~5 in the comment \cite{Condmat}}
Figure 5 in the comment \cite{Condmat} presents the temperature dependences of $\Im \rho^{1\omega}$ and ${\rm d}\rho_0/{\rm d}T$ for Y$_{1-x}$Tb$_{x}$Mn$_6$Sn$_6$ ($x=0, 0.07, 0.10$). At 500 Hz, $\Im \rho^{1\omega}$ is suppressed as the amount of Tb substitution increases, whereas ${\rm d}\rho_0/{\rm d}T$ is weakly dependent on substitution. With this observation, the authors \cite{Condmat} concluded that the Joule-heating-induced AC electrical response is not relevant to the nonlinear impedance observed in YMn$_6$Sn$_6$. In the original paper \cite{KitaoriPRB}, Kitaori \textit{et al}.~claimed that by Tb substitution, the pinning effect and, thus, the characteristic frequency for the magnetic texture dynamics would naturally increase, resulting in a decrease in $\Im \rho^{1\omega}$ measured at 500 Hz. However, it is difficult to clearly determine the origin of the observed systematic changes caused by Tb substitution from the available data \cite{Condmat, KitaoriPRB} for the following reasons.

\begin{figure*}
\includegraphics{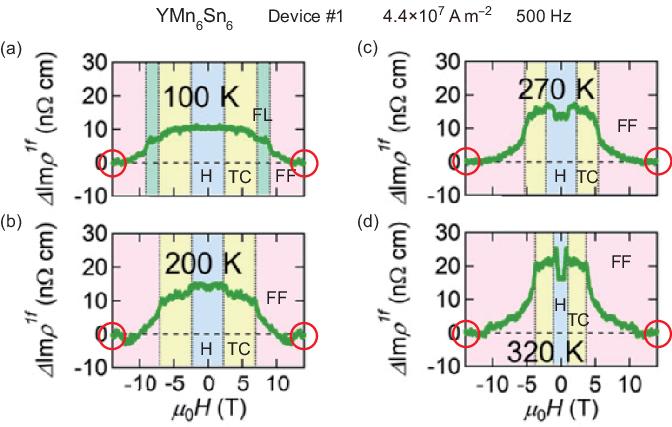}
\caption{\label{KitaoriVariousT} Magnetic field dependence of $\Delta \Im \rho^{1\omega}$ at selected temperatures for a low current excitation, $j_0 = 4.4$$\times$10$^7$ A~m$^{-2}$. The figure is reproduced from the literature \cite{KitaoriPNAS}. Kitaori \textit{et al.}~\cite{KitaoriPNAS} stated that the data were corrected by eliminating the background under the assumption that the signal of $\Im \rho^{1\omega}$ is zero (red circles) in the ferromagnetic phase. The labels H, TC, FL, and FF, denote the proper-screw helical, transverse conical, fan-like, and forced ferromagnetic phases, respectively. 
}
\end{figure*}

\begin{figure*}
\includegraphics{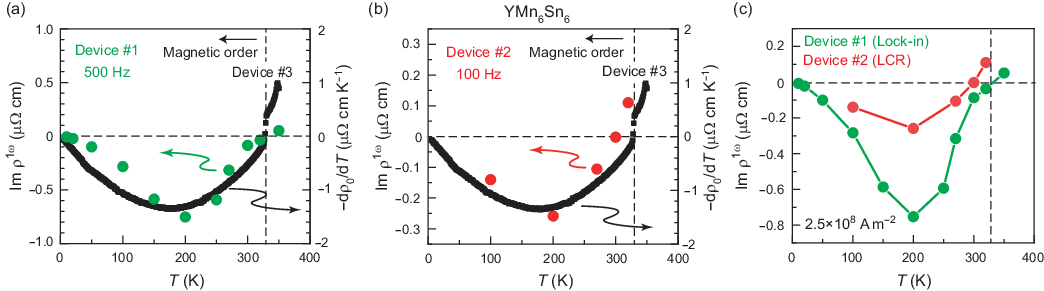}
\caption{\label{KitaoriCompare} Correlation between nonlinear $\Im \rho(j_0)$ and $-{\rm d}\rho_0/{\rm d} T$ in YMn$_6$Sn$_6$. (a) Comparison between $\Im \Delta \rho^{1\omega}(j_0)$--$T$ (device $\#$1) and the $\left( -{\rm d}\rho_0/{\rm d} T \right)$--$T$ profiles (device $\#$3). (b) Comparison between $\Im \Delta \rho^{1\omega}(j_0)$--$T$ (device $\#$2) and $\left( -{\rm d}\rho_0/{\rm d} T \right)$--$T$ profiles (device $\#$3). (c) Comparison of $\Im \Delta \rho^{1\omega}(j_0)$--$T$ profiles between devices $\#$1 and $\#$2. The $\Im \rho^{1\omega} (j_0)$--$T$ profiles were obtained from the literature \cite{KitaoriPNAS}. Device $\#3$ was measured using a lock-in amplifier, whereas device $\#2$ was measured using an LCR meter.
}
\end{figure*}

First, in the original paper on YMn$_6$Sn$_6$ \cite{KitaoriPNAS}, the authors explicitly stated that, ``The contribution from parasitic capacitance of the circuit becomes evident in the low-current region ($j_0 < 2$$\times$10$^8$ A~m$^{-2}$) with small signals of $\Im \rho^{1\omega}$. Such extrinsic contributions are subtracted under the assumption that the signal of $\Im \rho^{1\omega}$ would be zero in the ferromagnetic phase.''. In fact, when closely examining the $\Im \rho^{1\omega}$--$H$ profiles measured at various temperatures with a relatively low current density, 4.4$\times$10$^7$ A~m$^{-2}$ [for clarity, we reproduce Fig.~2J--M in Ref.~\cite{KitaoriPNAS} as Fig.~{\ref{KitaoriVariousT}}(a)--(d)], the value of $\Im \rho^{1\omega}$ always reaches zero at 14 T in the forced ferromagnetic (FF) phase for all temperatures. The method of analysis differs from that for Gd$_3$Ru$_4$Al$_{12}$, in which the nonlinear impedance remains nonzero in the ferromagnetic phase (Fig.~{\ref{Current_dep}}). Therefore, if all the data presented in the original papers by Kitaori \textit{et al}.~\cite{KitaoriPNAS, KitaoriPRB} were processed in the same way, the $\Im \rho^{1\omega}$ data may represent relative changes from the ferromagnetic phase at 14 T for each temperature. Since the data before background subtraction are not available in the literature \cite{KitaoriPNAS, KitaoriPRB}, it is difficult for us to provide a detailed discussion of Y$_{1-x}$Tb$_{x}$Mn$_6$Sn$_6$. We agree that the $\Im \rho^{1\omega}$ data are generally susceptible to background signals from the measurement circuit and therefore some data processing is needed to discuss the $\Im \rho^{1\omega}$ of a material under investigation. However, if some signals, due to spin fluctuations \cite{Fluctuations} and/or Joule heating \cite{PrepriFuruta}, appear in the ferromagnetic phase, the data processing approach mentioned by Kitaori \textit{et al}.~\cite{KitaoriPNAS} is inappropriate. As in the discussion on Gd$_3$Ru$_4$Al$_{12}$, the $\rho^{3\omega}$ data, which are less susceptible to parasitic components of the circuit, could avoid this issue. Therefore, the unprocessed $\Im \rho^{3\omega}$ would be more helpful than the processed or unprocessed $\Im \rho^{1\omega}$.

In the comment \cite{Condmat}, the authors argued that the temperature dependences of $\Im \rho^{1\omega}$ (device $\#$3) and ${\rm d}\rho_0/{\rm d}T$ (device $\#$1) are not tightly correlated with each other in YMn$_6$Sn$_6$. For clarity, we reproduce Fig.~{\ref{KitaoriCompare}}(a). Note that we do not know how the complex $\rho^{1\omega}$ measured using a lock-in amplifier were processed by referring to the $\Im \rho_0^{1\omega}$--$H$ profile. In contrast, in our preprint \cite{PrepriFuruta}, we compare the ${\rm d}\rho/{\rm d}T$--$T$ profile (device $\#$1) to the $\Im \rho^{1\omega}$--$T$ profile (device $\#$2) [reproduced as Fig.~{\ref{KitaoriCompare}}(b)], because the device $\#2$ is measured using an LCR meter only at 0 T. Therefore, it is unlikely that they refer to the data of the ferromagnetic phase at 14 T when background subtraction is performed. Thus, we speculate that the $\Im \rho^{1\omega}$--$T$ profiles obtained in device $\#$2 may be more reliable than those obtained with device $\#$1. Nevertheless, the $\Im \rho^{1\omega}$--$T$ profiles obtained from devices $\#$1 and $\#$2 do not coincide with each other, even though they are measured using the same current density, 2.5$\times$10$^8$ A~m$^{-2}$ [Fig.~{\ref{KitaoriCompare}}(c)]. In the context of the Joule-heating-based scenario, this observation implies that $\Im \rho^{1\omega}$ is sensitive to subtle issues, such as the data processing method, the amount of Joule heating, and the heat dissipation to the heat bath.

Second, for the substituted materials, no frequency-dependent data was reported either in the original paper \cite{KitaoriPRB} or in the comment \cite{Condmat}. Thus, it is still not clear whether the characteristic frequency truly increases upon Tb substitution. This issue still remains to be verified, as mentioned in the original paper \cite{KitaoriPRB}.

\begin{figure*}
\includegraphics{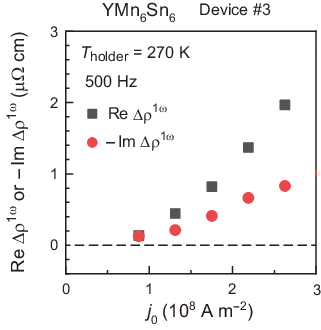}
\caption{\label{KitaoriCurrent} Current-density dependencies of complex $\Delta \rho^{1\omega} (j_0) = \rho^{1\omega} (j_0) - \rho^{1\omega} (0.44\times10^8 {\rm A~m^{-2}})$ at $T_{\rm holder} = 270$ K. This figure is constructed from raw data published in the literature \cite{KitaoriPNAS}.
}
\end{figure*}

Third, the information regarding the real part of the nonlinear impedance is not given for the substituted materials. As shown in Fig.~{\ref{KitaoriCurrent}}, the nonlinear impedance of YMn$_6$Sn$_6$ measured at 500 Hz, which is lower than the characteristic frequency of the material ($\approx$1 kHz) \cite{KitaoriPNAS}, shows that $\Re \Delta \rho^{1\omega} > \Im \Delta \rho^{1\omega}$ and is therefore also resistor-like in nature, as in the cases of Gd$_3$Ru$_4$Al$_{12}$ and FeSn$_2$ \cite{PrepriFuruta}. Thus, when considering the origin of the nonlinear impedance, the real part should be more informative than the imaginary part is. As demonstrated in Sections~III and VI C, focusing only on the imaginary part of resistive-like nonlinear impedance can lead to indirect arguments and an increased risk of misinterpretation.

Fourth, it is also important to verify the cross-sectional area dependence of the nonlinear impedance, as was done for Gd$_3$Ru$_4$Al$_{12}$ \cite{YokouchiNature} and FeSn$_2$ \cite{YokouchiArxiv}. In fact, for Gd$_3$Ru$_4$Al$_{12}$, the extensive dataset has been discussed in the light of the nonlinear EEF mechanism, and it was initially assumed to demonstrate $\Im Z^{1\omega} \propto S^{-3}$ behavior; however, the experiments eventually exhibit $\Im Z^{1\omega} \propto S^{-2}$ behavior (Fig.~{\ref{Nature_S_scaling}}), contradicting the initial assumptions. In addition, given the case study of Gd$_3$Ru$_4$Al$_{12}$, the in-plane-field angle dependence is also helpful to verify whether the observed nonlinear impedance is truely associated with the STT effect.

Finally, to check the relevance of the Joule-heating-induced AC electrical response by comparing different devices, the information regarding Joule heating and/or $\Delta T$ is necessary. Because the amount of Joule heating applied to each device varies with the contact resistance, which is related to the area of the electrodes, it is unclear whether comparing devices at a certain current density guarantees that the amount of Joule heating and/or $\Delta T$ is the same between different devices. Furthermore, as mentioned above, if the data are processed so that $\Im \rho^{1\omega}$ measured at relatively weak current densities always reaches zero at 14 T, perhaps the data of $\Im \rho^{1\omega}$ should be viewed as representing relative changes from the ferromagnetic phases in each material. The systematic change with Tb substitution is an interesting issue, but given that it is not easy to prepare the same conditions of Joule heating and thermal relaxation across devices, thorough measurements on each device, including $\rho^{3\omega}$ and frequency dependence, are also crucial.

For these reasons, we do not believe that Fig.~5 in the comment \cite{Condmat} present sufficient evidence to argue that the Joule-heating-induced AC electrical responses are not relevant to the data. In Ref.~\cite{KitaoriPNAS} and the comment \cite{Condmat}, the authors concluded that the current-induced EEF due to the magnetic textures dynamics is the origin of the nonlinear impedance. Since neither the cross-sectional area dependence nor the in-plane-field angle dependence of the nonlinear impedance has been clarified, the EEF-based scenario cannot be entirely ruled out, unlike the case of Gd$_3$Ru$_4$Al$_{12}$. However, the EEF-based scenario has thus far failed to predict even the approximate shape of the $\Im \rho^{1\omega}$--$H$ profile. In contrast, the Joule heating model can predict the $\Im \rho^{1\omega}$--$H$ profile given the ${\rm d}\rho_0/{\rm d}T$--$H$ profile. Therefore, it would be interesting to compare these two profiles. Unfortunately, the ${\rm d}\rho_0/{\rm d}T$--$H$ profile has not been clarified for YMn$_6$Sn$_6$.

\subsection{VII. Conclusion}
In the comment \cite{Condmat}, new data were presented. On the basis of these data, the authors claimed that the nonlinear impedance is due to the EEF mechanism. However, in Gd$_3$Ru$_4$Al$_{12}$, this claim is contradicted by the observation of $\Im Z^{1\omega} \propto S^{-2}$. Furthermore, the in-plane-field angle dependence of the nonlinear impedance also contradicts the scenario that the nonlinear impedance originates from the EEF induced via the STT effect. The authors \cite{Condmat} claimed that there are data that cannot be explained by our Joule heating model. However, this is because the authors misinterpreted the scope of the applicability of the $P_0$-linear minimal Joule heating model and because the data that the authors considered are under the influence of the extrinsic SC and singularities associated with a phase transition. Regarding Y$_{1-x}$Tb$_{x}$Mn$_6$Sn$_6$, the situation is much more unclear compared with that for Gd$_3$Ru$_4$Al$_{12}$ due to the lack of various fundamental information as well as some uncertainties regarding data handling.

We emphasize that the EEF-based scenario does not currently offer any detailed explanation regarding the temperature and magnetic-field dependencies of the nonlinear impedance and its order of magnitude. In fact, the $\Im \rho^{1\omega}$--$H$ and $\Im \rho^{1\omega}$--$T$ profiles observed in the experiments \cite{YokouchiNature, KitaoriPNAS, YokouchiArxiv} have never been reconstructed even at a qualitative level from the perspective of the EEF-based scenario. In contrast, these issues can be addressed, at least qualitatively, through the Joule-heating-based scenario \cite{PrepriFuruta}, provided that detailed data on the resistivity in the linear response are available. Finally, since the nonlinear impedance has been found to be resistor-like, the nonlinear impedance observed to date \cite{YokouchiNature, KitaoriPNAS, YokouchiArxiv} should be considered in terms of the nonlinear resistance measured at a finite frequency. Thus, analyzing the real part is more important than analyzing the imaginary part. A bias toward analyzing the imaginary part could lead to incorrect insights regarding the origin of the nonlinear impedance.



\begin{thebibliography}{99}
\bibitem{Note} Note that if the $S^{-3}$ scaling is observed, it would imply a bulk mechanism, but not necessarily a nonthermal one. Indeed, as discussed in the next paragraph, the Joule heating in the bulk can result in nonlinear $\Im Z^{1\omega} /d$ that scales with $S^{-3}$ at a given $I_0$.
\end{thebibliography}
\end{document}